\documentclass[11pt,a4paper]{article}

\usepackage{jheppub}
\usepackage{amsthm}

\usepackage{multirow, graphicx,amssymb,url,mathrsfs,amsmath}
\usepackage{wrapfig,boxedminipage,setspace,subfigure,epsfig}
\usepackage{amsxtra,amstext,latexsym,dsfont,amsfonts}
\usepackage{color,eucal}
\usepackage[dvipsnames]{xcolor}
\usepackage{float}
\usepackage{slashed,comment}
\usepackage{kotex}
\usepackage{tensor}

\usepackage{indentfirst}

\newtheorem{theorem}{Theorem}[section]

\newcommand{\ket}[1]{\mbox{$| #1 \rangle$}}

\newcommand{\Tr}{{\rm Tr}\,}

\newcommand{\ie}{{\it i.e.}}

\newcommand{\ee}{entanglement entropy}
\newcommand{\Ee}{Entanglement entropy}

\title{Generalized Rindler Wedge and Holographic Observer Concordance}

\author[a]{Xin-Xiang Ju,}
\author[b]{Wen-Bin Pan,}
\author[a,c]{Ya-Wen Sun}
\author[a]{and Yuan-Tai Wang}

\emailAdd{juxinxiang21@mails.ucas.ac.cn}
\emailAdd{panwb@ihep.ac.cn}
\emailAdd{yawen.sun@ucas.ac.cn}
\emailAdd{wangyuantai19@mails.ucas.ac.cn}

\affiliation[a]{School of Physical Sciences, University of Chinese Academy of Sciences, Zhongguancun east road 80, Beijing 100190, China}
\affiliation[b]{Institute of High Energy Physics, Chinese Academy of Sciences,19B Yuquan Road, Shijingshan District, Beijing 100049, China}
\affiliation[c]{Kavli Institute for Theoretical Sciences, University of Chinese Academy of Sciences, Beijing 100049, China}

\abstract{Defining gravitational subsystems has long been challenging due to the lack of the conventional notion of locality in gravity. In this work, we define gravitational subsystems from the observable spacetime subregions of a set of well-defined accelerating observers. 
We study the most general horizons of accelerating observers and find that in a general spacetime, only spacelike surfaces satisfying a global condition could become horizons of well-defined accelerating observers, which we name the Rindler-convexity condition. The entanglement entropy associated with a Rindler-convex region is proportional to the area of the enclosing surface. The subregions defined from this observer perspective is named the generalized Rindler wedge. This
provides a physical origin for defining gravitational subsystems associated with one type of Type III von Neumann subalgebra. 
 We propose the holographic interpretation of generalized Rindler wedges and provide evidence from the observer correspondence, the subregion subalgebra duality, and the equality of the entanglement entropy, respectively. 
We introduce time/space cutoffs in the bulk to substantiate this proposition, generalize it, and establish a holographic observer concordance framework, which asserts that the partitioning of degrees of freedom through observation is holographically concordant.}

\begin{document}
\maketitle

\tableofcontents
\section{Introduction} 
\noindent

The definition of independent subsystems in a gravitational spacetime has long posed a challenge due to the absence of the conventional notion of locality in such systems \cite{Giddings:2015lla,Pollak_2018}. In holography, based on the Ryu-Takayanagi formula \cite{Ryu:2006bv}, the entanglement wedge (EW) is a natural way to define a spacetime subregion whose degrees of freedom are fully encoded within this subregion \cite{Czech:2012bh,Wall:2012uf,Headrick:2014cta,Bousso:2022hlz,Espindola:2018ozt,Saraswat:2020zzf,dong2016reconstruction,Harlow:2018fse,Bousso:2012sj}. The subregion of spacetime within the entanglement wedge corresponds to a boundary subsystem, which is referred to as the subregion-subregion duality. The existence of islands in the entanglement wedge of Hawking radiations \cite{Almheiri_2020,Almheiri_2021} further reveals the nonlocality of gravitational degrees of freedom which cannot be divided in an arbitrary way.

Exploring the existence of more generalized subsystems in gravity and holography presents an intriguing and significant question \cite{Balasubramanian:2023dpj}. Subsystems in gravity could be consistently defined through the subalgebra associated with this subregion of spacetime \cite{Giddings:2018cjc}. In \cite{Leutheusser:2022bgi}, two different types of Type III von Neumann subalgebras are associated to spacetime subregions in a gravitational system. The first type corresponds to the subalgebra of entanglement wedges, while the second type could be defined to more general spacetime subregions, including the causal wedges. Subregion-subalgebra duality could be thought of as a kind of generalization of the subregion-subregion duality. In the perspective of defining gravitational subsystems, this also generalizes the usual entanglement wedge to more general spacetime subregions. In the construction of the subalgebra associated with the more general spacetime subregions, a geodesic convexity condition needs to be taken into account for the consistent definition of such a subalgebra.

In this work, we provide a physical origin for this subalgebra definition of the more generalized subsystem in general gravitational spacetime. We define well-defined gravitational subsystems using accelerating observers. The spacetime subregions that could be observed by a set of consistently defined accelerating observers could be well-defined subregions. These subregions are surrounded by the horizon of a set of accelerating observers. For the set of accelerating observers in this spacetime subregion to be well-defined, a convexity condition has to be imposed, reducing to the geodesic convexity condition in the case of \cite{Leutheusser:2022bgi}. Our perspective of accelerating observers provides a more general division of spacetime subregions, and is consistent with the subregion-subalgebra duality in \cite{Leutheusser:2022bgi}.

To do this, we start by constructing null hypersurfaces that could be horizons for accelerating observers from spacelike surfaces. We find that not all spacelike surfaces could become horizons for accelerating observers. It turns out that in general spacetimes, only surfaces obeying a global condition, namely the Rindler-convexity condition, could be horizons of well-defined observers. We name the subregion that could be observed by these accelerating observers the generalized Rindler wedges (GRW). 

Another motivation for studying the horizons for general accelerating observers comes from the microscopic interpretation of the black hole horizon entropy in gravity. Black holes have a Bekenstein-Hawking entropy proportional to the area of their horizons  \cite{Bekenstein:1973ur,Bardeen:1973gs,Bekenstein:1974ax,Hawking:1975vcx,Eisert:2008ur,Jacobson:2003wv}. It has long been a mystery as to what are the microscopic degrees of freedom composing the entropy of a black hole. 
One perspective on this question is that not only black hole horizons, but also observer horizons could
be associated with an entropy proportional to its area, e.g. the Rindler horizon  \cite{Rindler:1966zz,Laflamme:1987ec,Czech:2012be,Bousso:2012mh}. This leads to one possible explanation that the gravitational entropy of a subregion/a horizon counts the degrees of freedom \cite{VanRaamsdonk:2010pw} that an accelerating observer cannot observe.

In our setup, the entropy associated with the horizon of general accelerating observers is shown to be proportional to the area of the surface in the observer spacetime. This entropy should be the entanglement entropy (EE) between the subregion enclosed by the surface and its complement in the original spacetime. 
Thus the horizon entropy could be regarded as arising from the degrees of freedom that the observer cannot observe.  




In this paper, we give the Rindler convexity condition, which is a constraint on the shape of accelerating observer horizons, and study its various interesting consequences for gravitational entanglement structure, including the validity of nesting rules and the saturation of the subadditivity condition. Then an important question arises as to what is the dual interpretation of these gravitational physics in holography. We propose a spacetime subregion duality for the generalized Rindler wedges, i.e. a generalized Rindler wedge corresponds to a boundary state in the spacetime subregion of the intersection of the GRW with the boundary spacetime. More explicitly, by the boundary state in a certain boundary spacetime subregion, we refer to the boundary state of the corresponding spatial subregion with certain long-range entanglement structure removed and which long-range entanglement structure needs to be removed is fully determined by the causal structure of the boundary spacetime subregion, i.e. the entanglement between any two small subregions which cannot be causally connected in the spacetime subregion needs to be removed. We will provide evidence for this proposal from the observer correspondence, the subalgebra-subregion duality, from the equivalence of the entanglement entropy calculated for the state in the spacetime subregion with the entanglement entropy obtained from the GRW surface area and from a volume law in the case of very small time range for the boundary spacetime subregion. 

An immediate question is if the entanglement wedge could also be a generalized Rindler wedge. In general, the answer is no when the entanglement wedge is not a causal wedge. This means that for entanglement wedges that are not equivalent to causal wedges, we cannot define a set of nonintersecting observer worldlines within this entanglement wedge. In this case, the entanglement wedge cannot be an exact observable region of a set of well-defined observers. Thus the observers in the entanglement wedge are a set of non-physical observers in this sense. Then could we turn an entanglement wedge to a 
generalized Rindler wedge? We propose a solution to this question by introducing spacetime cutoffs in the gravitational spacetime. As special cases and further generalizations, we also study spacetimes whose IR geometries are modified or even removed, which correspond to boundary states with different long-range entanglement structures.

The rest of the paper is organized as follows. In section 2, we propose and explain in detail the Rindler-convexity condition and the generalized Rindler wedges in any gravitational spacetime. In section 3, we calculate the entanglement entropy of GRWs and show the validity of the nesting rules and subadditivity condition. Sections 2 and 3 concentrate on gravitational physics without incorporating the concept of holography. In section 4, we study the holographic boundary interpretation of this entanglement entropy and the boundary dual of the GRWs. In section 5, we study spacetime cutoffs in the gravitational spacetime and its interesting consequences. Section 6 is devoted to conclusions and discussions.

\section{The Rindler-convexity condition} 
\noindent
As we have previously discussed, gravitational degrees of freedom exhibit non-local characteristics \cite{Giddings:2015lla,Pollak_2018}; thus, an arbitrary partition of spatial regions may not correspond to physical degrees of freedom. In gravity, therefore, subsystems could be defined through the subalgebra associated with this subregion of spacetime \cite{Giddings:2018cjc}. In \cite{Leutheusser:2022bgi}, subalgebra for more general spacetime subregions other than the entanglement wedges has been given. However, a geodesic convexity condition is required in \cite{Leutheusser:2022bgi}. Here, we provide a physical perspective for this definition of the more generalized subsystem in general gravitational spacetime. We define well-defined gravitational subsystems using the spacetime subregions that could be observed by consistently defined accelerating observers. This requires the {spatial} subregion to be exactly the observable region of a set of well-defined observers \footnote{In this work, by observers we refer to a set of worldlines that fills the observable spacetime, i.e. at a fixed time slice, there would be one observer at each spatial point.}, i.e. the surface of this region should be the horizon of accelerating {observers}.


    The simplest and most well-known observer horizon is the Rindler horizon for a uniformly accelerating observer in flat or AdS spacetime. {In the Rindler wedge within Minkowski spacetime {or AdS}, it is important to note that at each {spatial point}, there exists a unique worldline, which corresponds to the worldline of the Rindler observer at rest in the Rindler frame. Therefore, there exists a set of infinitely many accelerating observers. {Each Rindler observer could have different accelerations}. 

        In  \cite{Balasubramanian:2013rqa,Huang:2007tw,Tian_2011} a sphere in flat or AdS spactime was shown to be the horizon for radially accelerating observers. A spherical Rindler coordinate transformation was employed to transform the surface of the sphere into a spherical Rindler horizon, i.e. the horizon of the set of radially accelerating observers.
    
    In more generalized cases, when considering a broader range of accelerating observers {with varying orientations of their 3-acceleration vectors,} it becomes evident that the shapes of their causal horizons exhibit greater diversity. We could start from an orientable spatial surface {emmbedded} on a Cauchy slice of a general spacetime and try to build null hypersurfaces from this space-like surface as causal horizons are null hypersurfaces. However, not all kinds of spacelike surfaces would {generate} horizons whose corresponding observers are physically well defined.

{Locally we could always perform a reference frame transformation to designate a specified null surface as the horizon of local accelerating observers. However, globally the worldlines of these accelerating observers could intersect, making them ill-defined.} Here we demonstrate that in a general spacetime, for a surface {embedded on a Cauchy slice} to be the horizon of an accelerating observer, the surface on the Cauchy slice must satisfy a global condition to avoid this intersection, which we refer to as the Rindler-convexity condition.

Note that what we are doing is to start from a flat spacetime or pure AdS spacetime or any given gravitational spacetime, and then pick any spacelike surface in this spacetime on a Cauchy slice to make it expand in both the future and the past directions at the speed of light therefore forming two null hypersurfaces.  We examine whether this could be associated with a set of well-defined accelerating observers. Those two null hypersurfaces serve as the past and future horizons of these observers. 
This Rindler-convexity condition is a constraint on the shape of this space-like surface at a fixed time slice in the original spacetime where the observers are accelerating. 

\begin{figure}[h]
    \centering 
    \includegraphics[width=0.6\textwidth]{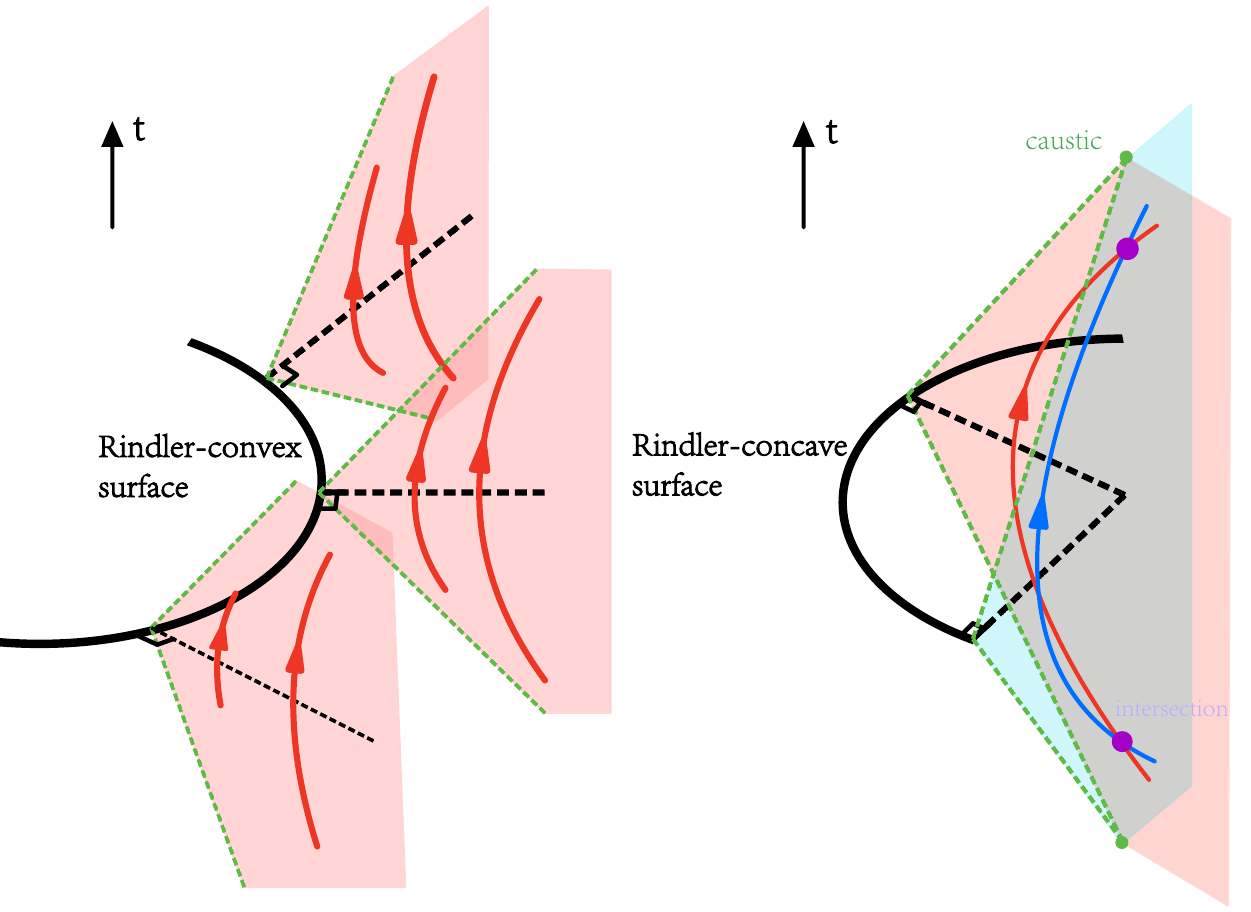} 
    \caption{Accelerating observers associated with Rindler convex and concave surfaces. The worldlines of observers are marked by arrowed curves (hyperbolas). The dashed green lines are {\it normal null geodesics} of the black space-like curves. It can be seen that the normal null geodesics and worldlines of the observers will not intersect when the surface is Rindler-convex (left), while intersect to form caustics (green dots) and the intersection point (purple dots) when the surface is concave (right).} 
    \label{convexconcave}
\end{figure}

{In Figure \ref{convexconcave}, we show an illustration of this spacelike surface and the corresponding worldlines of observers in a 2+1 dimensional spacetime. On the left, the spacelike surface is Rindler-convex, which expands towards outside along with time. Accelerating observers live outside this curve whose worldlines do not intersect. On the right, a concave curve is shown, whose accelerating observers outside cannot avoid intersections. This closed curve is a Rindler concave one according to our definition. In flat spacetime, Rindler convexity reduces to the usual convexity.}

Under this Rindler-convexity condition, the spacetime subregion that could be observed by these well defined accelerating observers are defined to be the generalized Rindler wedge (GRW). In this paper, we also name the spatial complementary region of the GRW as a Rindler convex region. In general, within the GRW, the 3-acceleration of these observers may not necessarily be aligned in the same spatial direction, different from the conventional Rindler observers.
    
We will give the definition and equivalent formulations of the Rindler-convexity condition in section 2.1. In section 2.2, we give some simple examples of GRWs.  In section 2.3, we will give the explicit geometric constructions of the GRW for some simple cases. We emphasize here that in this and the next section, we will concentrate on the gravitational physics without incorporating the concept of holography.



    \subsection{Derivation of the Rindler-convexity condition}

The condition for the null hypersurface to be a horizon comes from the fact that the worldlines of accelerating observers should not intersect with each other so that the reference frame transformation makes sense physically. The observer horizon is a null hypersurface and we can view it as a space-like surface expanding at the speed of light along the normal directions {(\ie its null generator)}  so that any {observer} behind the horizon can never catch the expansion of the horizon, no matter in which direction it moves. From this viewpoint, the nonintersecting condition imposes a constraint on the shape of the spacelike surface. We can analyze it as follows.

{For a given spacelike hypersurface {on a Cauchy slice} to become a bifurcation surface of the horizon of observers, there should exist at least a consistent set of observers at each spatial point {outside the horizon} who will not {cross} the horizon. In the critical case, the accelerating observer near the bifurcation surface must move in the spatial direction perpendicular to the surface to maximize their chance of avoiding falling into the hypersurface. Consequently, this bifurcation surface could generate a horizon for these observers. The worldlines of these observers are hyperbolas in the spacetime with normal null geodesics being their asymptotes. Illustrations of normal null geodesics are shown in Figure \ref{convexconcave} by green dashed lines. Thus this no intersection condition indicates that {\it the normal null geodesics of the causal horizon must never intersect.}} This leads to the following definition of Rindler-convexity from the normal null direction.

   \textbf{Normal condition for the definition of Rindler-convexity:}
    a region on a Cauchy slice is `Rindler-convex' if the normal null geodesics outside its boundary both to the future and from the past never intersect to form caustics.

 We have already obtained the definition of Rindler-convexity, which constrains the shape of the bifurcation surface of observer horizons. However, this definition is not conveniently used in telling if a given spacelike surface could be Rindler convex or not. Therefore, we develop the following equivalent definition of Rindler-convexity.

\textbf{Tangential condition for the definition of Rindler-convexity:}
    the boundary of a region on a Cauchy slice is `Rindler-convex' if any lightsphere externally tangential to its boundary\footnote{{It is known in mathematics that the supporting hyperplane is used to test the convexity of a set in the topological vector space $X=\mathbb{R}^n$. Here, we use lightspheres as the supporting hyperplane to test the Rindler convexity of a hypersurface on a Cauchy slice.}} never reaches the inside of the region, where a `lightsphere' is defined to be the intersection of the Cauchy slice with an arbitrary lightcone.

This definition is based on the following observation. The right side of Figure \ref{convexity} shows that why a concave surface (the surface of the green shaded region on the $t=0$ Cauchy slice) cannot serve as an observer horizon. If it does, the region on the $t=1$ Cauchy slice which causally connected with the green region is bounded by the green curve. This could be seen as the horizon expanding along the null generators marked by the blue arrows. Then one can found that an observer marked by the purple triangle in the red shaded region cannot escape from the expansion of the horizon no matter which direction it moves. 
    As a result, this concave surface cannot serve as the horizon for observers inside the red shaded region. This means that for concave surface $\partial A$, we cannot found a family of accelerating observers causally connected with the entire $A^c$ and causally disconnected with $A$ for arbitrarily long time. This leads to the following fact as well as the tangential formulation of the definition of Rindler-convexity. {\it Any point inside the `Rindler-convex hull' $B$ of a given region $A$ will inevitably have causal connection with region $A$.}
    `Rindler-convex hull' of A {is defined as} the smallest Rindler-convex region $B\supset A$.

    \begin{figure}[h]
        \centering 
        \includegraphics[width=1\textwidth]{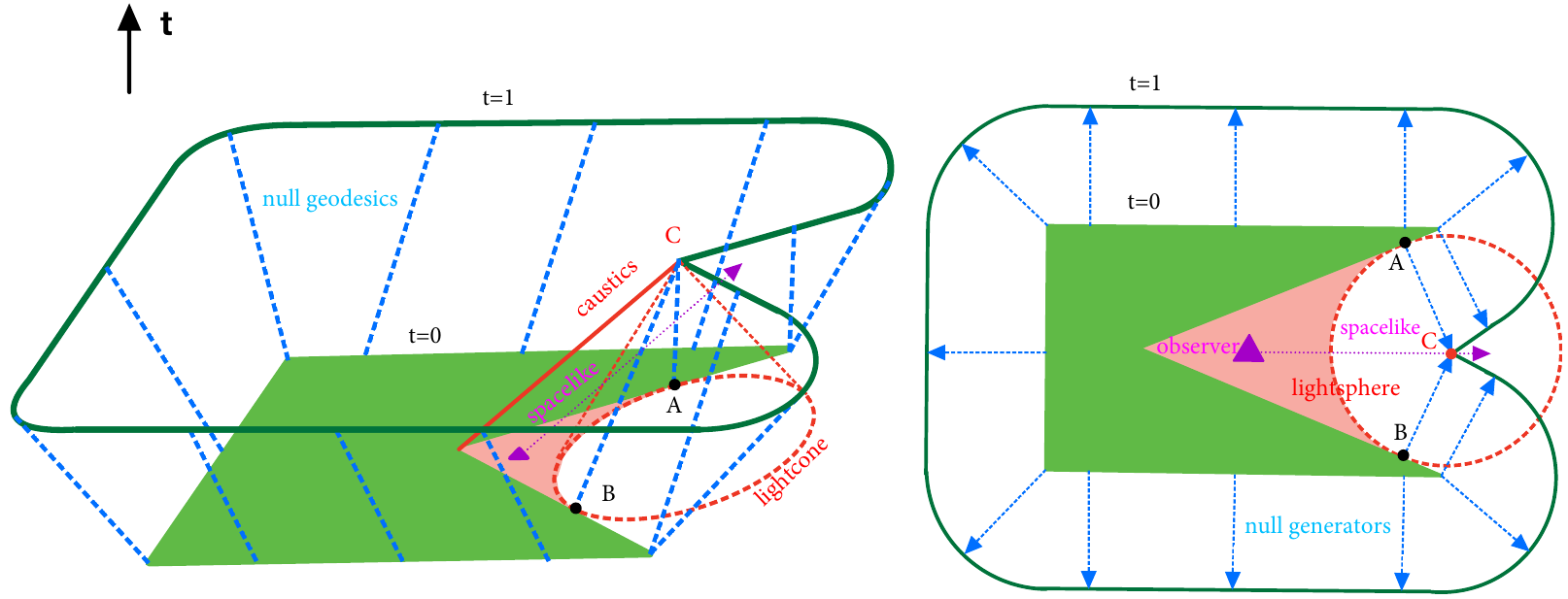} 
        \caption{A Rindler-concave spatial region (green) whose normal null geodesics intersect. The green concave region on the $t=0$ Cauchy slice is expanding along the null generators marked by the blue arrows (blue dashed lines on the left side). On the $t=1$ Cauchy slice, the region causally connected with the green region is bounded by the green curve. The red dashed circle represents a lightsphere externally tangential to the green region. The spatial distances $|AC|$ and $|BC|$ are both equal to 1. Any observer $O$ inside the red shaded region on the $t=0$ Cauchy slice has a spatial distance $|OC|>1$, making it impossible to reach the outside of the green curve on the $t=1$ Cauchy slice regardless of the direction it moves. This results in the surface of the green region being unable to serve as an observer horizon. The caustic on the $t=1$ Cauchy slice, formed by the intersection of the normal null geodesics $AC$ and $BC$, is marked by the red dot $C$ (red line on the left side).}
        \label{convexity}
    \end{figure}

    Then we have a natural guiding principle to define gravitational subsystems, i.e. degrees of freedom that {a family of} accelerating observers can observe. This demands that the {spatial} subregion has to obey the Rindler-convexity condition. When the whole original spacetime is in a pure state, the observable subregion would be in a {thermal-like} state with a finite temperature. In this scenario, the thermal entropy of the complement (i.e., the observable spacetime) of the Rindler-convex region in the accelerated reference frame should correspond to the entanglement entropy of the Rindler-convex region with its complement. This is analogous to the case of Rindler spacetime, where the thermal entropy of the radiation in Rindler is equivalent to the entanglement entropy of the Rindler wedge with the rest of the spacetime in the vacuum state.
    
    We refer to the causal domain of the complement of the Rindler-convex region $A$ on the Cauchy slice, the spacetime subregion where accelerating observers can access, as the generalized Rindler wedge, denoted as $D(A^c)$\footnote{{Note that the term “Rindler-convex region” refers to a spatial region on a Cauchy slice that is bounded by a Rindler-convex surface, whereas “GRW” denotes the spacetime subregion that is bounded by a null hypersurface.}}. It is worth mentioning that in sections 2 and 3, we concentrate on gravitational entropy, {the entanglement entropy between pure gravitational degrees of freedom (GRW and its complement)} as mentioned above, as distinct from the holographic entanglement entropy and the concept of differential entropy discussed in the holographic context in section 4.

Now we introduce two statements regarding the definition of Rindler convexity. The first is that the normal condition and the tangential condition of Rindler convexity \footnote{Note that Rindler-convexity is a covariant concept that does not rely on reference frames.  Moreover, as Weyl transformation leaves null geodesics invariant, Rindler-convexity is also Weyl invariant.} are equivalent. The second is that when $T_{\mu\nu}k^\mu k^\nu=C_{\rho\mu\nu\sigma}k^\rho k^\sigma=0$, on a Cauchy slice with zero extrinsic curvature $K_{\mu\nu}=0$, Rindler-convexity is equivalent to geodesic convexity. The second effect also explains the word ``convexity" in the definition of Rindler-convexity as in flat spacetime, this reduces to the usual meaning of convexity for surfaces.

Now we prove these two statements in order. Readers who are particularly interested in the physics aspects may skip this part and proceed directly to the next subsection.

\begin{theorem}
    \textit{{ The normal condition and the tangential condition of Rindler convexity \footnote{Note that Rindler-convexity is a covariant concept that does not rely on reference frames.  Moreover, as Weyl transformation leaves null geodesics invariant, Rindler-convexity is also Weyl invariant.} are equivalent.}}
\end{theorem}
\begin{proof}
     First, if the normal condition is violated for a hypersurface $\partial H$ on the Cauchy slice, i.e., the normal null geodesics intersect to form a caustic $C$ on the $t=t_c$ Cauchy slice. The lightsphere formed by the intersection of the light cone at $C$ and the $t=0$ Cauchy slice must be tangential to $\partial H$ at two points, $A$ and $B$ (Figure \ref{convexity}). Then, the lightspheres tangential to $\partial H$ at $A$ which are larger than it must reach the inside of the region $H$, and the tangential condition is violated. On the other hand, if the tangential condition is violated, there must exist a lightsphere $L$ tangential to $\partial H$ at two points. As a result, the normal null geodesics emitted from those two points must intersect to form a caustic $C$ where the light cone on it intersects with the $t=0$ Cauchy slice, forming the lightsphere $L$, and the normal condition is violated. To summarize, the normal condition is equivalent to the tangential condition.
\end{proof}

    Here we provide some remarks on the question of how ``general" the Rindler-convex condition is. Firstly, any Killing horizon in a static spacetime is Rindler-convex because its normal null vectors are Killing vectors that never intersect. On a flat Cauchy slice in flat spacetime, it is evident that Rindler-convex regions are equivalent to geometric convex sets (hence the term ``Rindler-convex"). The equivalence between Rindler-convexity and geodesic convexity in the most trivial case is not coincidental. {In fact, we have:}
    
    \begin{theorem}
        \textit{When $T_{\mu\nu}k^\mu k^\nu=C_{\rho\mu\nu\sigma}k^\rho k^\sigma=0$, on a Cauchy slice with zero extrinsic curvature $K_{\mu\nu}=0$, Rindler-convexity is equivalent to geodesic convexity, where $T_{\mu\nu}$ is the energy-momentum tensor, $C_{\rho\mu\nu\sigma}$ is the Weyl tensor, and $k^\mu$ is any null vector. }
    \end{theorem}
    \begin{proof}
        The Proof can be found in appendix A. 
    \end{proof}
    Spacetimes satisfying $T_{\mu\nu}k^\mu k^\nu=C_{\rho\mu\nu\sigma}k^\rho k^\sigma=0$, which we refer to as ``null vacuum," include maximally symmetric spacetimes such as de Sitter (dS) and anti-de Sitter (AdS) vacuum geometries.
    
   {However, in the presence of matter that deforms the spacetime according to the Einstein equation, null geodesics tend to converge (and intersect) due to gravity, {as described by Raychaudhuri's equation.}
   \begin{equation}
        \frac{d\theta}{d\lambda}=-\frac1{D-2} \theta^2-\hat\sigma_{\mu\nu}\hat\sigma^{\mu\nu}-8\pi T_{\mu\nu}k^\mu k^\nu.
    \end{equation}
   This leads to a stricter notion of Rindler-convexity. Specifically, when a black hole is present, any subregion outside the black hole is Rindler-concave because the normal null geodesics must converge and ultimately intersect when they pass near the event horizon, due to the gravitational lensing effects. Therefore, a Rindler-convex surface must wrap around the event horizon.}

In a recent paper \cite{Leutheusser:2022bgi}, it is shown that on a time-reflection symmetric Cauchy slice in AdS, only the causal domain of geodesic convex subregions could be described holographically by the commutant of the algebra of a time-band on the boundary. As the extrinsic curvature is zero on this Cauchy slice, geodesic convexity equals Rindler-convexity in their case. In \cite{Hubeny_2014} a condition was also proposed to avoid the multivalue problem of the transferring function in hole-orgraphy, which is equal to the Rindler-convexity condition {in the scenario of holographic bulk spacetime.} These results further confirm the special role that Rindler-convex regions play in related physics and it would be interesting to know if geodesic convexity in  \cite{Leutheusser:2022bgi} could be generalized into Rindler-convexity in more general situations. The application of the subregion subalgebra correspondence in GRWs will also be further discussed in section 4 of this paper.

    \subsection{Examples of Rindler convex surfaces and GRWs}
   {In order to prevent the above discussion from being overly abstract, here we provide some simple and specific examples of GRWs and Rindler convex surfaces on a Cauchy slice with zero extrinsic curvature in various background geometries as follows.}
   \begin{itemize}

       \item {\it A Rindler wedge in AdS or in Minkowski spacetime is a GRW}. {In Rindler wedge, Rindler transformation guarantees that a set of consistent accelerating observers exist in the Rindler wedge with the Rindler horizon being their horizon.} By the definition of Rindler-convexity, a Rindler wedge is a GRW.

       \item {\it In Minkowski spacetime, on a flat Cauchy slice, surface $\partial A$ is Rindler-convex if and only if $A$ is a convex set. }As we have proved, in flat spacetime, Rindler convexity reduces to the usual notion of convexity. Therefore in Minkowski spacetime, any convex spatial surface could be the bifurcation surface of the horizon of well-defined accelerating observers.

       \item {\it In de-Sitter vacuum, the static patch is the smallest GRW with the cosmological horizon being its Rindler-convex surface, which has the largest surface area}, as described in Figure \ref{dssphere}.
       \item {\it A spherical spatial region \cite{Balasubramanian:2013rqa} outside the trapped/anti-trapped surface in a spacetime with spherical symmetry is Rindler-convex}, as described in Figure \ref{dssphere}.
       \item {\it The intersection of a killing horizon with a Cauchy slice with zero extrinsic curvature is Rindler-convex.} {As there are no caustics on a killing horizon, thus normal condition of Rindler convexity is satisfied.}
       
       \item {\it In AdS vacuum or a BTZ black hole geometry, the causal wedge ($CW(A)$) of a boundary subregion $A$ is a GRW, while its entanglement wedge ($EW(A)$) is generally not a GRW (Section 4).}
   \end{itemize}

\begin{figure}[h]
    \centering 
    \includegraphics[width=0.7\textwidth]{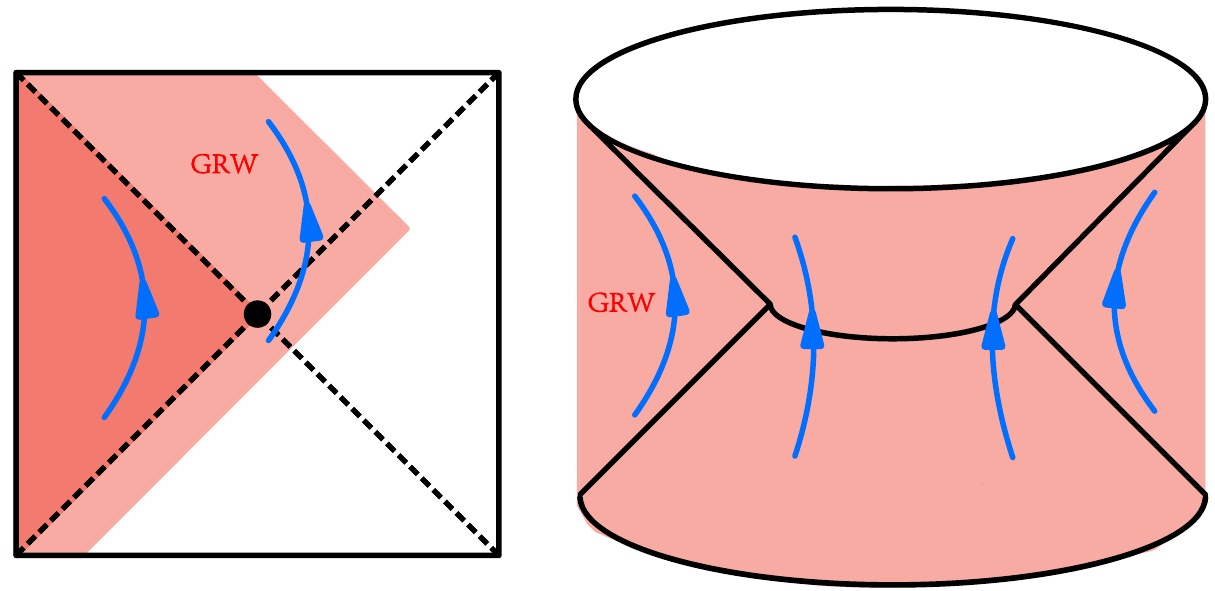} 
    \caption{Left: in dS spacetime, the static patch is the smallest GRW with the largest surface area, which is the cosmological horizon. A larger GRW is also plotted as the light red shaded region in the figure, which has a smaller surface area, though.  Right: in a spherically symmetric spacetime, the radially radiating null geodesics will never intersect if it is not inside an trapped/anti-trapped surface. Thus, a spacial sphere is always Rindler-convex under this circumstance.}
    \label{dssphere}
\end{figure}

Besides these simple examples, we could also produce Rindler convex regions through the following procedure.
\begin{theorem}\label{union}
    \textit{If $A$ and $B$ are Rindler convex regions on a Cauchy slice, and $A\cap B$ has a smooth boundary, then $A\cap B$ is also a Rindler convex region.}
\end{theorem}
 \begin{proof}
     {Given $C=A\cap B$, we have:}
    \begin{equation}
        \partial C=(\partial C \cap \partial A )\cup(\partial C \cap \partial B )
    \end{equation}
    because of the Rindler-convexity of regions $A$ and $B$, and due to the tangential condition, the lightspheres tangential to $\partial A$ ($\partial B$) will never reach the inside of $A$ ($B$). As a result, the lightsphere tangential to $\partial C$ that is either tangential to $A$ or $B$ will never reach the inside of $A\cap B$, which makes $C$ a Rindler-convex region by the tangential condition.
 \end{proof}

    Note that any smooth Rindler-concave surface could also look like a horizon locally by a coordinate transformation, however, as we have shown explicitly, there cannot exist globally well-defined accelerating observers for concave subregions as the worldlines of the accelerating observers would intersect at a spacetime point indicating that the observer at that point cannot be uniquely defined, thus the degrees of freedom in a concave subregion cannot be separated from its complement in a consistent way. This Rindler-convexity condition is a {\it global} condition and this may also imply that the entropy is also related to the global structure of the horizon. 
    
    \subsection{Explicit geometric constructions of Rindler-convex regions in various spacetime}
    Given a specific Cauchy slice in a specific spacetime, one would like to {to obtain explicitly geometry constructions of Rindler-convex surfaces and corresponding generalized Rindler wedges.} In principle, we can find all the lightspheres on this Cauchy slice and then use the tangential condition to determine the Rindler-convexity condition. However, as a larger lightsphere always can contain a smaller one due to causality, the tangential condition for a larger lightsphere is always stronger than the tangential condition for a smaller lightsphere. As a result, we only have to use the infinitely large lightsphere emitting from (converging to) past (future) null infinity to test the Rindler convexity.

In the following, we construct explicit Rindler convex surfaces in three types of geometries, the $2+1$ dimensional conformally flat spacetime with the flat spacetime as a special case, the $2+1$ dimensional conformally flat spacetime with conformal boundaries with pure AdS being a special case, and the four dimensional Schwarzschild black hole spacetime.

}

\noindent{\bf Case I}: a conformally flat spacetime without a spacetime boundary in the conformal coordinate.

The spacetime metric is:
\begin{equation}
ds^2=\Omega^2(t,x,y)(-dt^2+dx^2+dy^2),\quad\quad (t,x,y \in \mathbb{R})
\end{equation}
where $\Omega^2(t,x,y)$ is a smooth positive function without singular values. {This includes the flat spacetime as a special case.} This spacetime could be viewed as a Weyl transformation acting on a 3D Minkowski spacetime. As a Weyl transformation preserves null geodesics \cite{Carroll:2004st}, null geodesics are \textit{straight lines} in conformal coordinates. On the $t=0$ Cauchy slice, lightspheres are circles, and the infinitely large lightspheres are straight lines. Due to the tangential condition, Rindler-convexity regions on the $t=0$ Cauchy slice are equal to the convex sets in the 2D flat Euclidean geometry of $x$ and $y$. We parametrize the Rindler convex surface, i.e. the boundary of a (strictly) convex region $A$, denoted as $\partial A$, to be $(x=x_0(\theta), y=y_0(\theta))$. The Rindler convexity condition requires that they satisfy the parametric equations as follows
\begin{equation}\label{parametric}
        \left\{
        \begin{aligned}
        x & = x_0(\theta)  \\
        y & = y_0(\theta) 
        \end{aligned}
        \right.,\quad\text{where} \quad \frac {dx_0(\theta)}{d\theta}\cos\theta +\frac {dy_0(\theta)}{d\theta}\sin\theta=0,\quad(\theta \in [0,2\pi)).
    \end{equation}
In this equation, $\theta$ denotes the azimuth of the vector normal to the convex curve. The convexity is reflected in the fact that $x_0(\theta)$ and $y_0(\theta)$ are single-valued functions.

The procedure to obtain any Rindler convex surface from this equation is the follows. We could choose an arbitrary function of $x(\theta)$ which is periodic in $theta$ and then obtain the corresponding $y(\theta)$ by solving the differential equation in (\ref{parametric}). In this way, all Rindler convex surfaces could be produced in this spacetime. The simplest example would be to choose $x(\theta)= \cos \theta$ and $y(\theta)$ could be solved to be $c+\sin\theta$, where $c$ is an arbitrary integration constant. This gives a circle in the conformally flat spacetime. We can check the shape of the surface is always convex by choosing arbitrary periodic functions of $x(\theta)$.

   Let us prove the convexity of $\partial A$ using both tangential condition and normal condition as follows. The equation of the infinitely large lightsphere tangential to $\partial A$ at $\theta_0$ is
\begin{equation}\label{support}
    y_0'(\theta_0)(x-x_0(\theta_0))-x_0'(\theta_0)(y-y_0(\theta_0))=0.
\end{equation}
Using equation (\ref{parametric}), one can prove that 
\begin{equation}
    y_0'(\theta_0)(x_0(\theta)-x_0(\theta_0))-x_0'(\theta_0)(y_0(\theta)-y_0(\theta_0))\leq0,
\end{equation}
where the equality holds if and only if $\theta=\theta_0$, which means that every point on $\partial A$ is on the same side of the infinitely large lightsphere.  In other words, line (\ref{support}) serves as the supporting hyperplane of the convex region $A$. Consequently, $\partial A$ is a Rindler-convex surface due to the tangential condition.

Testing Rindler-convexity via the normal condition is straightforward. In conformal coordinates, the null geodesics normal to $\partial A$ at $(0,x_0(\theta_0),y_0(\theta_0))$ are given by
\begin{equation}
     \left\{
        \begin{aligned}
        t & = t_0 \quad (t>0)\\
        x & = x_0(\theta) + t_0\cos\theta\\
        y & = y_0(\theta) + t_0\sin\theta 
        \end{aligned}
    \right. \quad\text{and}\quad
    \left\{
        \begin{aligned}
        t & = t_0 \quad (t<0)\\
        x & = x_0(\theta) - t_0\cos\theta\\
        y & = y_0(\theta) - t_0\sin\theta 
        \end{aligned}
    \right..
\end{equation}
These geodesics point to the future and the past, respectively. Let us choose two arbitrary null geodesics pointing to the future (without loss of generality) with $\theta=\theta_1$ and $\theta=\theta_2$ respectively. One can calculate the distance between two points $d(t_0)$ on these geodesics with the same $t_0$. This distance is a monotonically increasing function of $t_0$, which means $d(t_0)\geq d(0)>0$. In other words, the normal null geodesics never intersect, verifying the normal condition.

\noindent{\bf Case II}: conformally flat spacetimes with conformal boundaries.

Conformal boundaries can generally exist in conformally flat spacetimes. The most well-known case is AdS spacetime,  in Poincaré half-plane coordinates, the metric given by:
\begin{equation}
ds_1^2=\frac1{z^2}(-dt^2+dx^2+dz^2),\quad\quad (t,x \in \mathbb{R}; z>0)
\end{equation}
  Here, the conformal boundary is located at $z=0$. 
As we argued in case I that null geodesics are still straight lines in conformal coordinates, the light-spheres will still be circles. However, the ``infinitely large lightsphere" is affected by the presence of the conformal boundary, which in turn, affects Rindler-convexity. Specifically, the ``infinitely large lightsphere" would be generated by the light cone whose vertex is on the conformal boundary.
From the perspective of the normal condition, the existence of a spacetime boundary relaxes the Rindler-convexity condition by making the null geodesics more difficult to intersect. Note that normal condition and tangential condition are still equivalent.

\begin{figure}[H]\label{PoincaréRindlerconvex}
    \centering 
    \includegraphics[width=0.95\textwidth]{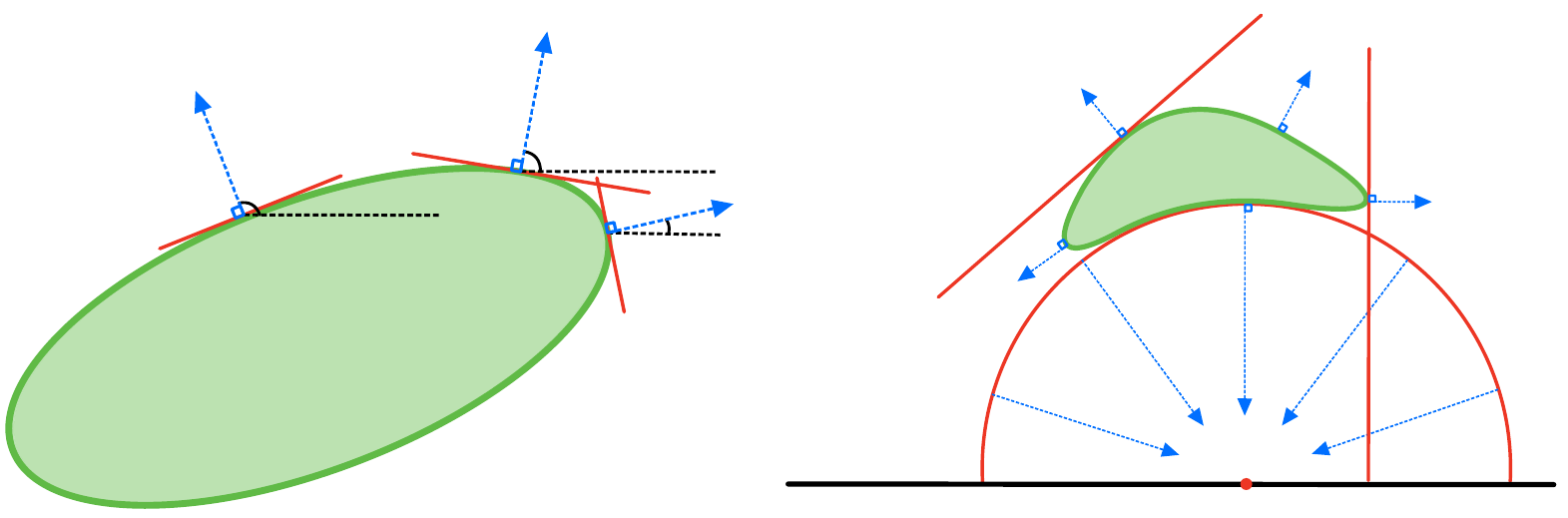} 
    \caption{Examples of Rindler-convex regions (green shaded) are shown on a flat Cauchy slice in flat spacetime (left) and in the AdS Poincaré patch (right). Blue arrows represent null geodesics, and red lines (curves) depict infinitely large lightspheres.
    } \label{examples}
\end{figure}

As shown in Figure \ref{examples}, with a conformal boundary existing, the infinitely large lightsphere is represented by the red semicircle. Coincidentally, it forms a geodesic in AdS spacetime, implying that Rindler convexity is equivalent to geodesic convexity in global AdS spacetime (Appendix A). 
{If we make the same ansatz as in the previous case, where we replace $y$ in (\ref{parametric}) by the $z$ coordinate, we can find that the surface could be Rindler convex even if $x_0(\theta)$ and $y_0(\theta)$ are not single-valued functions in certain cases. An example is shown as the right side of Figure \ref{PoincaréRindlerconvex}. In order to get the universal form of Rindler-convex surfaces, we should use another coordinate $\lambda$ to reparameterize $x_0$ and $z_0$ functions as follows.}
\begin{equation}
    \left\{
    \begin{aligned}
    x & = x_0(\lambda)  \\
    z & = z_0(\lambda) 
    \end{aligned}
    \right.,\,\,\text{where}\,\, 
    \left\{
    \begin{aligned}
    &\lambda=\theta\,\, \text{for}\,\, {\theta}\in (0,\pi)  \\
    &\tan \lambda=x_0(\theta)-z_0(\theta)\cot \theta,\,\,\text{for}\,\,\theta \in [\pi,2\pi]. 
    \end{aligned}
    \right.
\end{equation}
Then, the surface could be Rindler-convex if and only if $x_0$ and $z_0$ are single-valued functions of $\lambda$, where $\tan \lambda$ is the $x$ coordinate of the intersection point between the normal null geodesic and the conformal boundary.

Moreover, it is worth noting that the Rindler-convexity in Poincaré coordinate is not equivalent to the Rindler-convexity in global AdS coordinate. The later is equivalent to geodesic convexity but the former is not. 
As shown in Figure \ref{Poincarépatch}, the red surface is Rindler-convex in Poincare coordinate but Rindler-concave in global AdS. 
This disparity arises because the Poincaré patch in global AdS spacetime has an additional spacetime boundary at $z=\infty$ and two null hypersurfaces, which relax the conditions for Rindler convexity.

    \noindent{\bf Case III:}{ Rindler-convex regions in Schwarzschild black hole spacetime.}

    We construct Rindler-convex surfaces in the case of a four-dimensional Schwarzschild black hole in flat spacetime. 
    As previously mentioned, the infinitely large lightsphere is used to test the Rindler-convexity in curved spacetimes.
    To obtain the trajectories of infinitely large lightspheres, in the following we compute the null geodesics emanated from a point source far away from the Schwarzschild black hole.
    Only the light ray propagating in the equatorial plane is needed because of the rotational symmetry around the line connecting the black hole and the source.
    The metric of the equatorial plane is
    \begin{equation}
    ds^2=-f(r)dt^2+\frac{1}{f(r)}dr^2+r^2d\phi^2,  \;\; f(r)=1-2M/r
    \end{equation}
    A null geodesic $x^\mu(s)$ parameterized by affine time $s$ obeys
    \begin{equation}\label{LS0}
    g_{\mu \nu}\frac{dx^\mu}{ds}\frac{dx^\nu}{ds}=-f(r)\left(\frac{dt}{ds}\right)^2+\frac{1}{f(r)}\left(\frac{dr}{ds}\right)^2+r^2\left(\frac{d\phi}{ds}\right)^2=0.
    \end{equation}
    Two conserved quantities, energy $E$ and angular momentum $L$, are associated to the photon traveling along the geodesic with
    \begin{equation}\label{LS1}
    E=f(r)\frac{dt}{ds},  \;\;\;\; L=r^2 \frac{d\phi}{ds}.
    \end{equation}
    Thus the null geodesic equation \eqref{LS0} has the form
    \begin{equation}\label{LS2}
    -E^2+f(r)\left(\frac{dr}{ds}\right)^2+L^2 \frac{f(r)}{r^2}=0.
    \end{equation}

    Near the light source positioned at $r_0$ which is far away from the black hole ($r_0\to\infty$ and $f(r_0)\to1$), it is found that
    \begin{equation}
    -E^2+p_r^2+p_\bot^2=0,
    \end{equation}
    where $p_r=\frac{dr}{ds}$ is the radial component of momentum at infinity and $p_\bot=\frac{L}{r_0}$ can be identified as the component of momentum in the $\phi$ direction which is perpendicular to the radial direction.
    The ratio of $p_\bot$ and $p_r$ represents the direction of photon emission and is related to the impact parameter $\lambda=\frac{L}{E}$ as 
    \begin{equation}
    \frac{p_\bot}{p_r}=\frac{\lambda^2}{r_0^2-\lambda^2}.
    \end{equation}

\begin{figure}[H]
        \centering 
        \includegraphics[width=0.8\textwidth]{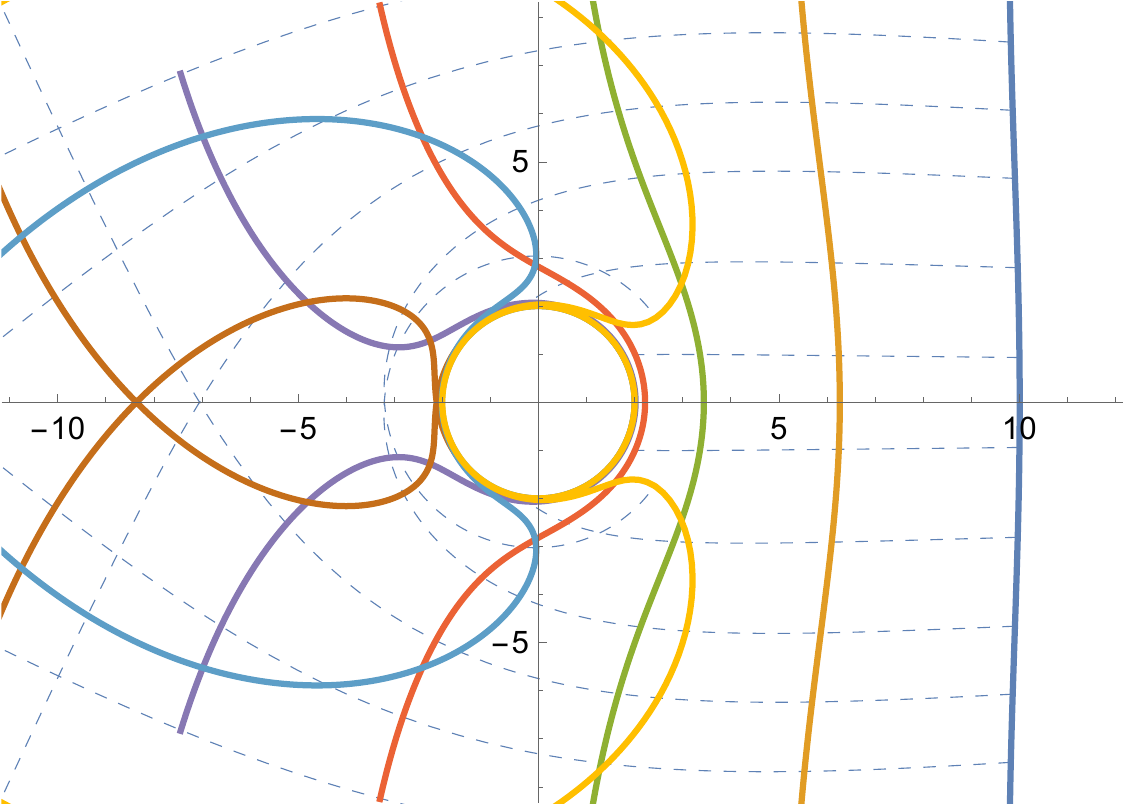} 
        \caption{Infinitely large lightspheres are depicted as colorful curves in four-dimensional Schwarzschild black hole spacetime. Regions on the right side of the lightspheres are generalized Rindler wedges. When lightspheres self-intersect, only the near-horizon regions surrounded by them are Rindler-convex.
        } \label{LS_Sch}
    \end{figure}

    Therefore, $\lambda$ labels the direction of a light ray emanated from a fixed distant source. The trajectories of photons and lightspheres in a Cauchy slice of time $t$ could be computed using \eqref{LS1} and \eqref{LS2}.
    For our purpose, we use the coordinate $t$ (instead of $s$) to indicate the time a photon travels along a specific geodesic and $\lambda$ to specify which geodesic it is on.
    Combining \eqref{LS1} and \eqref{LS2}, the set of equations determining $r(t,\lambda)$ and $\phi(t,\lambda)$ is found to be 
    \begin{equation}
    \left \{ \begin{matrix}
    \frac{d\phi}{dt}=\lambda \frac{f(r)}{r^2} \\
    \left(\frac{dr}{dt}\right)^2=f^2(r)\left(1-\lambda^2\frac{f(r)}{r^2}\right)  
    \end{matrix} \right. 
    \end{equation}
    and the initial condition is $r(0,\lambda)=r_0$ for arbirary $\lambda$.
    We solve this set of equations numerically.
    The photon trajectory of impact parameter $\lambda*$ is then given by $(r(t,\lambda*),\phi(t,\lambda*)$ with $0<t<\infty$, whereas the lightsphere at time $t*$ is given by $(r(t*,\lambda),\phi(t*,\lambda))$ with $-\infty<\lambda<\infty$.
    The photon trajectories and lightspheres of a distant source are shown in Figure \ref{LS_Sch}.

After we have obtained the explicit solutions of photon trajectories and infinitely large lightspheres, we have already found a set of Rindler-convex surfaces, which are these infinitely large lightspheres, as shown in Figure \ref{LS_Sch}. The GRWs are the subregions on the side of these lightspheres without the black hole horizon. Besides these Rindler convex surfaces, using theorem \ref{union}, we can choose the intersection of the infinite lightspheres to generate more Rindler convex surfaces and GRWs. 

\section{\Ee{} of GRW} 
\noindent
Generally, for a Rindler-convex surface on a Cauchy slice,{accelerating observers can always be constructed outside the surface, with the Rindler convex surface serving as the bifurcation surface of their horizon.} As a result, a coordinate transformation can always be found {in principle} to transform to the observer spacetime, {\ie, the GRW}. To calculate the entanglement entropy of this Rindler-convex region, we can then calculate the gravitational entropy in the observer spacetime, which is equal to the entanglement entropy of the Rindler-convex region in the original spacetime {as we take the whole spacetime as a pure state}. 

Note that {this procedure to calculate the entanglement entropy follows and generalizes \cite{Balasubramanian:2013rqa} for the ``hole" in flat spacetime, while this procedure is also similar to what \cite{Casini:2011kv} has done in the bulk for calculating the entanglement entropy of a spherical subregion at the boundary. In \cite{Casini:2011kv}, a bulk isometry transforms the RT surface of the boundary spherical region to a horizon of a topological black hole with a hyperbolic horizon, which is in fact a hyperbolic surface in AdS. After the transformation, the entanglement entropy in the original spacetime becomes the thermal entropy in the spacetime with a horizon. The same happens here in our system. After transforming to the observer spacetime, the thermal entropy is the same as the entanglement entropy in the original spacetime. As we will see in the next section, in fact \cite{Casini:2011kv} corresponds to the case that EW=GRW, i.e. the entanglement wedge of the boundary spherical region is a GRW, one special case here. We will calculate this entanglement entropy using the gravitational path integral method, the same as in \cite{Balasubramanian:2013rqa}. A calculation of the modular Hamiltonian needed in the holographic construction of \cite{Casini:2011kv} will be presented in another work \cite{Ju:2025}.}

Here as a simple example, we explicitly show a generalized Rindler coordinate transformation in a $2+1d$ conformally flat spacetime which has $ \left\{\begin{aligned}x & = x_0(\theta)\\y & = y_0(\theta)\end{aligned}\right. $ as the parametric function of a Rindler-convex horizon on a 2d Cauchy slice, where $\theta$ denotes the azimuth of the normal null geodesic. 
    The transformation is 
   \begin{equation}\label{GRT}
        \left\{
        \begin{aligned}
        x & = x_0(\theta)+\rho \cosh\tau\cos\theta,   \\
        y & = y_0(\theta)+\rho \cosh\tau\sin\theta,   \\
        t & = \rho\sinh\tau,
        \end{aligned}
        \right.
    \end{equation}
    where $ \rho$ and $\tau$ are the other two new observer coordinates. The Rindler-convexity of the horizon is revealed by the fact that $x_0(\theta)$ and $y_0(\theta)$ are both single-valued functions.
    The original metric is $ds^2={A(t,x,y)}^2(-dt^2+dx^2+dy^2)$. We get the metric of the observer spacetime 
    \begin{equation}\label{metric}
        ds^2=A^2\bigg(-\rho^2d\tau^2+d\rho^2+[{(\frac{dx_0(\theta)}{d\theta})}^2+{(\frac{dy_0(\theta)}{d\theta})}^2+2\rho \frac{\cosh\tau}{\cos \theta}\frac{dy_0(\theta)}{d\theta}+\rho^2\cosh^2\tau]d\theta^2\bigg).
    \end{equation} The horizon of the new spacetime is at $\rho=0$, at which $g_{tt}=0$ indicating that it is the surface of infinite redshift. As the surface gravity is a constant, there are no dynamics (energy, entropy flows) passing through the horizon so GRW is in thermal equilibrium. All cross terms are eliminated by the normal condition: $\cos \theta\frac{dx_0(\theta)}{d\theta}+\sin\theta\frac{dy_0(\theta)}{d\theta} =0$, {that is, on the Cauchy slice, the spatial trajectories of the observers are perpendicular to the bifurcation surface. That is, the worldlines of near-horizon observers should be on the timelike hypersurface normal to the bifurcation surface, shown as the shaded plane in Figure \ref{convexconcave}.} Because of this condition we do not need to worry about any divergence of the metric at $\theta=0, \pi/2$ as $y'(\theta)$ should cancel the zero of $cos(\theta)$ at $\theta=\pi/2$. 
    The temperature of this gravitational spacetime is $\frac{1}{2\pi\rho}$. Following  \cite{Balasubramanian:2013rqa}, we calculate the gravitational entropy for ($\ref{metric}$) using the replica trick and the result is
    \begin{equation}
        S=\frac{\mathcal{A}}{4G}，
    \end{equation}
    where $\mathcal{A}$ is the area of the horizon. Details can be found in appendix B.

\subsection{Inequalities of \ee{} {of generalized Rindler wedges}}
Let us now examine the consequences of the Rindler convexity condition on the consistent definition of subregions, leading to the validity of the following two inequalities: the nesting rule and the strong subadditivity inequality.

\noindent{\it The nesting rule.}
   If we do not impose this Rinder convexity condition on the shape of a well-defined subsystem in gravity, we would encounter a question: { what would happen if the boundary of the region becomes so ``rough" that its area becomes enormously large? This could lead to an unacceptable scenario where the gravitational entropy in some finite volume region becomes nearly infinite. In our framework, only subregions of generalized Rindler wedges or entanglement wedges can be well-defined gravitational subsystems. Importantly, Rindler-convexity prevents this paradox by disallowing the boundary of the region from becoming excessively ``rough".}
    
    {Let us delve into this topic more specifically. Firstly, to link degrees of freedom with entanglement entropy, we consider the Page Theorem} \cite{Page:1993df}, which suggests that when the dimensions of two Hilbert spaces are significantly different, the two subsystems are almost maximally entangled if the whole system is in a pure state:
    \begin{equation}
        S_A=-\Tr(\rho_A \log \rho_A ) \sim \log N,\quad \rho_A=\frac{I_N}{N},
    \end{equation}
    where $N$ is the dimension of the Hilbert space. The maximally entangled state has the maximum \ee{} which is exactly the coarse gained entropy, {\ie{} its degrees of freedom}  \cite{Harlow:2014yka,Almheiri:2020cfm}. 
    Here the Page theorem should be applicable to the space divided into two subregions with one of them having finite volume and the other having infinite volume. Thus we can always approximate the entanglement entropy of the finite volume region as $\log N$.

    Rindler-convexity guarantees that the surface area of a Rindler-convex region $A$ is larger than its Rindler-convex subregion under the null energy condition (NEC), \ie{}, $T_{\mu\nu}k^\mu k^\nu\geq0$ for any null vector $k^\mu$. We will refer to this inequality as the `surface area condition'  from now on. The proof can be found in appendix C. As a result, a larger volume has more degrees of freedom and the entanglement entropy.
    
    Surface area condition gives an inequality of the entanglement entropy of Rindler-convex regions under NEC
    \begin{equation}
        S(\rho_A)> S(\rho_B) \quad\quad (\text{A,B are both Rindler-convex and }A\supset B)
    \end{equation}
    We will refer to this inequality as the nesting rule (terminology borrowed from \cite{Bousso:2022hlz}). If concave regions could also have well-defined entanglement entropy proportional to their surface area, this nesting rule would be violated and this further confirms the important role of Rindler-convexity in the physical consistency of the system. 

    {It's noteworthy that when a matter field is introduced, by applying the generalized second law for causal horizons \cite{Wall_2013} and the Bousso bound \cite{Bousso:1999cb} instead of Raychaudhuri's equation, one can derive the following nesting rule:}
    \begin{equation}
        S_{gen}(A)> S_{gen}(B) \quad\quad (\text{A,B are both Rindler-convex and }A\supset B),
    \end{equation}
    where the gravitational entanglement entropy is replaced by the generalized entropy \cite{Bekenstein:1974ax}.
       
\noindent{\it Strong subadditivity.} 
    The strong subadditivity of the entanglement entropy in the gravitational spacetime requires
    \begin{equation}\label{SSA}
        S(\rho_{A})+S(\rho_{B})\geq S(\rho_{A\cup B})+S(\rho_{A\cap B}),
    \end{equation} 
    where $A$, $B$ are {subregions on a Cauchy slice}. 
    When $A\cap B=\varnothing$, the strong subadditivity inequality reduces to the subadditivity inequality. In our framework, well-defined entanglement entropy only exists for Rindler-convex regions. When $A, B$ are both Rindler-convex compact sets, their intersection, $A\cap B$ must be a Rindler-convex set, so we need to further assume that $A, B$, and $A\cup B$ are all Rindler-convex, i.e. we focus on these special cases. As we have
    \begin{equation}\label{ASSA}
        Area(A)+Area(B)= Area(A\cup B)+Area(A\cap B),
    \end{equation}
    where $Area(M)$ represents the surface area of the spatial subregion M. With entanglement entropy proportional to the corresponding area, we find that the entanglement entropy defined in this way for Rindler-convex regions naturally saturates the strong subadditivity condition. This saturation property leads to more profound physics which we report in \cite{Ju:2023dzo}. There is one subtle point here that in the proof for the strong subadditivity condition in quantum information theory \cite{Lieb:1973cp}, the entanglement entropy of region $A-A\cap B$ and $B-A\cap B$ should both be well-defined, which in our case may not be true, however, we still have the strong subadditivity condition obeyed in our case.

{\subsection{Connection with Bousso bound and other related concepts}}\noindent
  The Bousso covariant entropy bound \cite{Bousso:1999xy} states that for a spacelike surface B with area A, if the expansion of the congruence is non-positive at every point on $L$, with $L$ the hypersurface generated by the null congruence orthogonal to B, then the matter entropy is bounded by $S \leq \frac{A}{4 G}$. The difference from our framework is that here we focus on the gravitational entropy while not the matter entropy. In addition, Rindler-convexity demands that the outgoing null expansion $\theta$ {never reaches negative infinity}, while the Bousso bound demands the ingoing expansion is non-positive. Under NEC, the Rindler-convexity is a stricter condition and a region satisfying the Bousso bound condition is not necessarily Rindler-convex.

Moreover, motivated by the second law of thermodynamics and Bousso bound, another constraint on the light-sheet called `preferred holographic screen' was proposed \cite{Bousso:1999cb,Bousso:2015mqa,Bousso:2015qqa,Engelhardt:2015dta}, whose area bounds the entropy of matter on any light-sheet orthogonal to B. `Preferred holographic screen' demands that the expansion of the normal null geodesic on it cannot change its signature, which is equivalent to Rindler-convexity (\ie{} the boundary of any Rindler-convex region must be the intersection of these future and past light-sheets) when and only when NEC holds and the whole spacetime is geodesic complete. Moreover, Rindler-convexity is Weyl invariant while the preferred holographic screen is not. {\footnote{Note that the spacetime subregion duality of the generalized Rindler wedge is different from the surface/state correspondence \cite{Miyaji:2015yva} in the following aspects. The surface state correspondence was based on the construction of MERA.  As we mentioned, they have coarse-grained the system and eliminated all short-range entanglement structures by coarse-graining and the distangler. In this case, when we both consider a closed surface in the bulk, in the GRW setup, a mixed state is considered while a pure state is considered in the surface/state correspondence.}}
\vspace{0.6cm}
\section{Holographic dual of GRW} \noindent
{The primary objective of this section is to examine GRW in the holographic context with the aim of establishing its dual interpretation on the boundary. To achieve this goal, we begin by comparing GRW with the causal wedge and entanglement wedge in the first subsection. Then, in the second subsection, we introduce a concept called GRW spacetime subregion duality, utilizing observer physics and the differential entropy formula proposed in \cite{Balasubramanian:2013lsa}. Finally, in the third subsection, we present supporting evidence for this proposal within the context of von Neumann algebra and quantum information theory.}

When the spacetime is asymptotic AdS, an immediate question arises regarding the boundary interpretation of this gravitational physics. As a natural way to define gravitational subregions, the Rindler-convex subregion could span the whole boundary spatial region (the complementary region of a closed subregion in the bulk) or only cover a subregion of the boundary spatial region. {For the former case with a closed subregion removed in the bulk, one interpretation of the entanglement entropy of the subregion could be to reflect the UV/IR entanglement in the boundary \cite{Balasubramanian:2011wt,Balasubramanian:2013lsa,Balasubramanian:2018uus,Engelhardt:2018kzk}. Note that here we further demand the `hole' in \cite{Balasubramanian:2013lsa} be enclosed by a Rindler-convex surface so that the subsystem can be well-defined \footnote{We also notice that similar constraints were proposed \cite{Engelhardt:2015dta} in a different framework motivated by the holographic screen, which in this global AdS case, is equivalent to Rindler-convexity.}.

There have been a lot of discussions on the boundary dual interpretation of GRWs complementary to closed subregions in hole-orgraphy. It has been shown that the area of the bulk closed subregion gives the dual differential entropy  \cite{Balasubramanian:2013lsa}. A physical explanation of differential entropy could be the cost of state merging of the sum of the subsystems \cite{Czech_2015}. However, it is still a problem as to what is the dual state that the complementary region of the closed subregion describes and if differential entropy could be a physical entropy of a certain state \cite{Balasubramanian:2018uus}. One possibility is that the dual state is the original dual state with long-range entanglement structure removed due to the bulk removal of the corresponding IR region. This leads to a problem if the corresponding state is a physical one or not, i.e if the long-range entanglement structures could be removed consistently or not.  

\subsection{GRW vs CW and EW}\noindent
GRWs extend all the way to the boundary and intersect with the boundary at a boundary spacetime subregion in a way similar to the entanglement wedge and causal wedge. 

\begin{figure}[H]
        \centering 
        \includegraphics[width=0.3\textwidth]{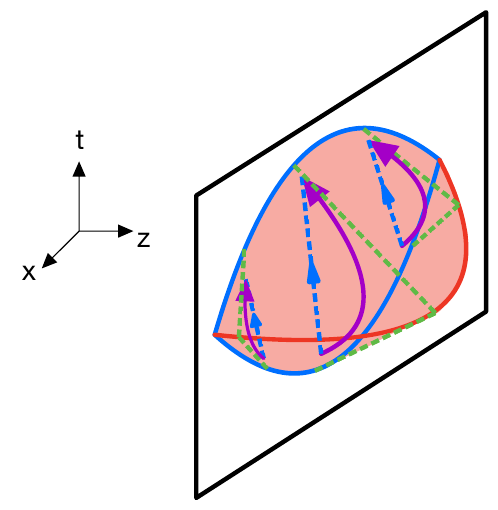} 
        \caption{A GRW (red shaded) and its intersection with the AdS boundary, i.e. the ``time pancake" bounded by the blue curves. The worldlines of Rindler observers in GRW are marked by curved purple arrows. Every bulk observer has a correponding homologous boundary observer, marked by the blue dashed line.
        } \label{GRWred}
    \end{figure}
Typically, accelerating observers live within this GRW subregion. An illustration of a GRW that intersects at the boundary with a spatial subregion is shown in Fig.\ref{GRWred}.  
Different GRWs could intersect the boundary with the same spatial boundary subregion while having different time ranges. {For the same boundary spacetime subregion, i.e. specifying both the spatial and time range, there is only one corresponding GRW extending to the bulk. The closer the surface of the GRW is to the boundary, the shorter the time range of the boundary spacetime subregion.}
For the same spatial subregion on the boundary, besides infinitely many GRWs, it also has corresponding entanglement wedge \cite{Czech:2012bh}, and causal wedge (CW) in the bulk. These different wedges might have different time ranges at the boundary. Here we point out the difference and connections between GRW, entanglement wedge and causal wedge for the same boundary spatial subregion, as well as their respective corresponding boundary degrees of freedom with a proposal for the boundary dual of GRW.

    {\it GRW vs CW\@.} {We denote the causal wedge (domains) for a finite boundary subregion $A$ as $CW(A)$.} {It is bounded by future and past causal horizon without caustics in the bulk in AdS vacuum background \cite{Hubeny_2013}\footnote{In the AdS Schwarzschild black hole geometry, caustics may form, and $CW(A)$ might exhibit non-trivial topology \cite{Hubeny_2013}. In such cases, $CW(A)$ does not qualify as a GRW.}.} Thus by the normal condition of Rindler-convexity, $CW(A)$ (in AdS vacuum) must be a GRW. 
    There are an infinite number of GRWs for the same boundary spatial subregion and the causal wedge is the outmost GRW among them as the causal wedge corresponds to the critical case where the normal null geodesics enclosing CW intersect at the AdS boundary \cite{Hubeny:2012ry}. An illustration can be found in Figure \ref{GRWUVIR}. {For the case of excited states? when $CW(A)$ has nontrivial topology \cite{Hubeny_2013}, GRW must be smaller than $CW(A)$ as it still has trivial topology.} {For more generally defined causal wedges, such as the ones causally connected to a spacetime subregion on the boundary, known as the ``strip wedge", it is not necessarily a GRW. Detailed discussions on this topic can be found in \cite{Hubeny_2014}. In this paper, we only refer to this $CW(A)$ as the causal wedge.}
    
    {\it GRW vs EW.} The relation between EW and CW has been discussed a lot in previous literature  \cite{Headrick:2014cta,Hubeny:2012wa,DeClerck:2019mkx}. In general, EW is deeper than CW in the bulk, except in certain special cases when EW=CW. 
    As any minimal surface is geodesic-concave except the planar ones, it can be proved that EW is bigger than GRW except when $\partial A$ is a $d-2$ sphere on the boundary so that EW=CW is a GRW. Particularly, in $AdS_3/CFT_2$, any connected region $A$ on the boundary has $\partial A$ being two points (0d spheres), so the EW of any connected region in $CFT_2$ has EW=CW being a GRW in the bulk.
   
 \begin{figure}[H]
        \centering 
        \includegraphics[width=0.6\textwidth]{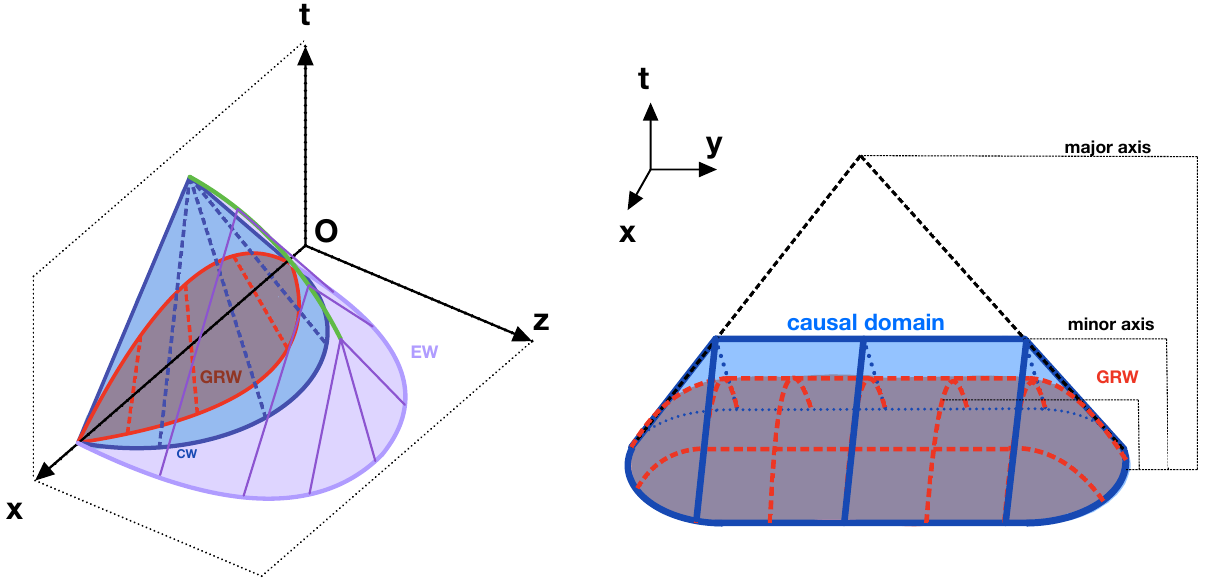} 
        \caption{GRW (red), CW (blue), and EW (purple) homologous to the same oval subregion on the boundary. Left: GRW, CW, and EW extending in the bulk radial direction $z$ and the boundary span in the time direction at $t>0$ shown in Poincar\'e coordinate with y-axis (major axis) compressed. $z$ is the bulk radial direction with $z=0$ the boundary. Right: the spacetime intersection of CW, GRW with the boundary. The height (time direction) of CW is equal to the minor axis while the height of GRW is less.} \label{GRWUVIR}
    \end{figure}

\subsection{The proposal for the dual state of GRW}

\noindent For the same boundary spatial subregion $A$, we have known that EW encodes all the information of $A$ \cite{Bousso:2022hlz,Espindola:2018ozt,Saraswat:2020zzf,dong2016reconstruction,Harlow:2018fse,Bousso:2012sj} in the sense that the area of EW counts the entanglement entropy of the boundary subsystem $A$ and the entanglement wedge could be fully reconstructed from the boundary operators in $A$. The corresponding causal wedge has been believed not to correspond to a known entanglement entropy when it is different from EW but encodes the uncertainty left after the observers measure all one-point functions inside the causal domain \cite{Kelly:2013aja}. Here we propose new boundary interpretations of GRW including CW as its special case based on the following facts.

I. From the geometry in the bulk, the depths of these wedges in the bulk are different. EW is deeper than CW in the bulk, and CW is deeper than general GRWs (Figure \ref{GRWUVIR}). This indicates that CW and GRW correspond to fewer IR boundary degrees of freedom compared to EW from the UV/IR correspondence, i.e. EW contains extra IR information compared to CW and GRWs.   
Here, combined with the holographic MERA correspondence\cite{swingle2012constructing,Vidal_2007,vidal2010entanglement}, which states that the radial degrees of freedom in gravity reflects the entanglement structure of the boundary theory at the corresponding scale, we interpret these IR degrees of freedom or IR information as encoding the long-range entanglement structure of the dual boundary spatial subregion. Thus the state that corresponds to the GRW should have less information in the long-range entanglement structure compared to the state corresponding to EW.
{On the other hand, the UV part of both the outermost GRW (CW) and EW of regular boundary subregions behaves similarly, as the UV part of both is always perpendicular to the asymptotic boundary. This inspires us to consider that the dual state of the GRW should exhibit the same short-range entanglement structures as the state dual to the EW.}

II. From the geometry on the boundary,
as the time dimension is included, the boundary spacetime subregions for different wedges with the same boundary spatial subregion become different in general (when the boundary spatial subregion is not a sphere), 
i.e. they cover different ranges in the time direction for the same spatial subregion $A$. We take the case when the boundary spatial subregion $A$ is an oval with a major axis $a$ and a minor axis $b$ as an example. In Figure \ref{GRWUVIR}), we can see that among all GRWs, CW covers the largest range in the time direction at the boundary, which is determined by the minor axis $b$, while GRW's range $\leq$ CW's. For the infinitely many GRWs, the closer the GRW is to the boundary, the shorter the time range is. For a given boundary spacetime subregion, the corresponding GRW is unique if it exists. This was also shown in \cite{Hubeny_2014} to come from the Rindler convexity condition.

III. From the observer perspective, the set of bulk accelerating observers induce a set of boundary observers in the intersecting spacetime subregion. The induced boundary observers might not have worldlines lying completely inside the boundary and it is possible to be a set of observers whose worldlines only exist for an arbitrarily small amount of time interval at a certain spacetime point in the boundary spacetime subregion. Even in cases when the boundary observers have their worldlines completely lying within the boundary, the boundary observers do not need to be accelerating, and in extreme cases, they could be staying at rest for their whole lifetime. A possible boundary spatial subregion, a time pancake shape, is shown in Figure \ref{GRWred}. This time pancake spacetime subregion is a causal incomplete subregion in the boundary spacetime and we can think of the observers as staying at rest at each spatial point while all having a birth and death time. Thus in this pancake subregion, for two small subregions with a spatial interval larger than the time range, no observer could have causal connections with both of these two small subregions at $t=0$. Thus in the observer language, no single observer could observe the two small subregions together. This means that the boundary observer cannot observe the entanglement structure between these two small subregions. Thus this time cutoff of boundary subregions of GRWs, i.e. their shorter time range, leads to less information on the long-range entanglement structure, which is consistent with larger entanglement entropy of GRWs. This is also consistent with the fact I, i.e. GRWs contain less information about the long-range entanglement structure from the bulk point of view. One illustration of this is shown in Fig. \ref{bcutoff}, where the time cutoff in the boundary leads to the ignorance of the entanglement between small boundary subregions $A$ and $B$ by the boundary observer, thus the effect is to cut off the entanglement between A and B and in the observer's perspective, A and B are both entangled with external degrees of freedom outside this spatial subregion. This leads to the growth in the entanglement entropy. From this, we see that the area of the GRWs, i.e. the bulk entanglement entropy between GRW and its complement still corresponds to an entanglement entropy at the boundary, but it is the entanglement entropy of a state that is not the real boundary state but a reconstructed state whose local density matrices are the same with the boundary state while some global entanglement structure is removed. 

More precisely, for the simplified shape of the time pancake of Fig.\ref{bcutoff}, we see that spatial subregion $A$ (or $B$, $E$) can be all observed by a single observer so the entanglement structure in $A$ (or $B$, $E$) can be fully observed, while A and B cannot be both observed by a single observer at the same time, so the boundary state observed by the induced boundary observers should be a state whose density matrices $\rho_{A}$, $\rho_{B}$, $\rho_{E}$, $\rho_{AE}$ and $\rho_{BE}$ are the same as the boundary state while $\rho_{AB}$ and $\rho_{ABE}$ are different with the boundary state as all entanglement between A and C are removed. 

Thus based on the facts above, we have the following proposal for the boundary dual of GRWs. {\it The boundary state dual to the bulk GRW is the state in the boundary spatial subregion with certain long-range entanglement removed and the causal structure of the boundary spacetime subregion determines which long-range entanglement should be removed, i.e. the entanglement between any two or multi subsystems which cannot be causally connected to a single boundary observer at the same time should be removed. }

From this point of view, the entanglement wedge describes all the degrees of freedom in ABE of Fig. \ref{bcutoff}, with all entanglement structures in $\rho_{ABE}$ kept, i.e. the EW describes the exact corresponding $\rho_{ABE}$ of the boundary spatial subregion $ABE$. In comparison, general GRWs describe a reconstructed density matrix of $\tilde \rho_{ABE}$ of the boundary spatial subregion $ABE$, with some long-range entanglement structure of $ABE$ removed. When the GRW goes closer to the boundary, the boundary intersecting subregion has a shorter time range, thus the dual state has more long-range entanglement removed with a larger fine-grained entanglement entropy. When the GRW is extremely close to the boundary, the time range in the boundary subregion is extremely short and the dual boundary state has almost all entanglement structure removed so that the boundary state is almost a product state of all the degrees of freedom in the spatial region. This is consistent with the fact that the area of GRW gives a volume law in this limit. More details will be discussed later. When EW is not a CW, EW corresponds to a nonphysical observer in the bulk as the enclosing surface is not Rindler convex. The observer who has access to the whole entanglement wedge would always observe a larger region than EW. Accordingly, the corresponding boundary observer of the EW is also not a physical one in the sense that when the observer observes all the degrees of freedom in the spatial subregion $A$, he/she has already observed extra degrees of freedom outside of $A$. Thus EW corresponds not to a physical observer but to an ``intelligent" observer who could analyze the data to figure out the degrees of freedom and the entanglement entropy of the subregion $A$ with information of a larger range of degrees of freedom. 

Note that here the dual state of GRW is not the real state of the dual field theory but it could be regarded as the ``real" state that the boundary observer observes. In this sense, the reconstructed state dual to the GRW should be a well-defined state with the long-range entanglement removed consistently by defining a set of physical observers. However, not all kinds of long-range entanglement structures can be removed consistently from a given state. This is the concern also in \cite{Czech_2015}, in which it was shown that the maximal entanglement entropy of the reconstructed state whose local density matrices have to be the same as the original state is the differential entropy, also equal to the area of the GRW surface. However, counterexamples exist showing that the maximal entropy of the reconstructed states in certain systems could not reach the differential entropy thus indicating that the state with long-range entanglement removed as required by the spacetime subregion causal structure might not be consistent. We will show in the next subsection that for holographic states this should not happen and we propose that long-range entanglement structures determined from the boundary spacetime subregion dual to GRWs can always be removed consistently for holographic states.

This proposal also provides a new explanation of the boundary interpretation of CW when CW is not equal to EW. It has long been a question as to what the causal information corresponds to when CW is not EW. One explanation is that it might correspond to a coarse-grained entropy as the area of a CW is larger than that of EW. Here we provide a new explanation that CW still corresponds to a fine-grained entropy but not of the original state but of the state with certain long-range entanglement structures removed.

 \subsection{Supporting evidence for the proposal} \noindent
 In the last subsection, we proposed the GRW/spacetime subregion duality hinted from the observation of the boundary observer together with the holographic MERA correspondence. We need further supporting evidence to show that this is correct. One concern is if the dual state with certain long-range entanglement structures removed is a consistent state. From the observer's perspective, as long as the bulk and boundary observers are consistently defined, the dual state should be a consistent one in the sense that it could be observed by a physical observer. Though counterexamples exist whose long-range entanglement could not be removed consistently, it is expected that for holographic states, this could always be done for the cases of consistent bulk and boundary observers. In this subsection, we will show further evidence for this duality, both from the subalgebra-subregion duality and the calculation of the entanglement entropy of the reconstructed boundary state whose long-range entanglement structure is removed in a certain pattern.

\subsubsection{Subalgebra-Subregion duality}
\noindent
For a holographic theory, a classical gravity theory is  equivalent to a boundary field theory and consequently the operator algebras of the two sides need to be the same. Subregion-subregion dualities then could be associated with a subregion-subalgebra duality \cite{Leutheusser:2022bgi}, i.e. a causally complete local bulk spacetime region has a one to one correspondence with an emergent type $III_1$ Von Neumann boundary subalgebra. 

The type $III_1$ boundary subalgebra in a spatial subregion $R$ associated with the bulk entanglement wedge is defined by $\mathcal{X}_R=\pi_{\Psi}(\lim_{N\to \infty,\Psi}\mathcal{P}_R \mathcal{B}^{(N)})$, where $\mathcal{B}^{(N)}$ is the boundary operator algebra at any $N$, $\mathcal{P}_R$ is the restricting operation to the region $R$ and $\pi_\Psi$ is the representation of the algebra in the Gelfand-Naimark-Sega (GNS) Hilbert space $H_{\Psi}^{GNS}$ associated with the semiclassical state $\ket{\Psi} $ following the conventions in \cite{Leutheusser:2022bgi}.

This subalgebra 
$\mathcal{X}_R$ has the property that for each Cauchy slice in the boundary subregion having the same causal completion $\hat{R}$, the subalgebra is the same so that this subalgebra could be considered to be associated with the boundary spatial subregion $R$. However, here for our purpose, as for the same boundary spatial subregion $R$ we have infinitely many different GRWs, we require a subalgebra that depends on the whole boundary spacetime subregion while not only the spatial subregion. This subalgebra is another type of sublagebra as already pointed out in \cite{Leutheusser:2022bgi}, defined as  $\mathcal{Y}_{\hat{R}}=\mathcal{P}_{\hat{R}}\pi_{\Psi}(\lim_{N\to\infty,\Psi} \mathcal{B^{(N)}})=(\mathcal{M}_{\hat{R}})''$, where $\hat{R}$ is the spacetime subregion of the causal completion of $R$ and $\mathcal{M}_{\hat{R}}$ is the restriction of the single-trace operator algebra $\mathcal{M}_{\Psi}$ on $\hat{R}$ and $'$ denotes its completion under the weak operator topology. Compared with the $\mathcal{X}_R$ subalgebra defined above, we could see that the difference between the two subregions is in the order of the $N\to \infty$ limit and the restriction to $\hat{R}$. These two operations do not commute in general, thus the two subalgebras are different in general. The algebra $\mathcal{B}^{(N)}$ has the property $\mathcal{B}_R^{(N)}=\mathcal{B}_{\hat{R}}^{(N)}$ for finite N, which is the reason that $\mathcal{X}_R=\mathcal{X}_{\hat{R}}$, which only depends on the spatial subregion $R$. However, taking first the large N limit would destroy this property. The $\mathcal{Y}_{\hat{R}}$ algebra depends on $\hat{R}$ and cannot be associated with only the spatial subregion $R$. Also $\mathcal{B}_{R}^{(N)}$ obeys additivity for finite N, i.e. $\mathcal{B}_{R_1}^{(N)} \vee \mathcal{B}_{R_2}^{(N)}=\mathcal{B}_{R_1\cup R_2}^{(N)}$. However, $\mathcal{X}_R$ has additivity anomaly due to the large N limit, i.e. $\mathcal{X}_{R_1}^{(N)} \vee \mathcal{X}_{R_2}^{(N)}\subseteq \mathcal{X}_{R_1\cup R_2}^{(N)}$.

This $\mathcal{Y}_{\hat{R}}$ is the subalgebra for causal wedges and the hole spacetime  corresponding to a boundary time band. These two systems are both related to GRWs and we could generalize the $\mathcal{Y}_{\hat{R}}$ algebra to obtain the corresponding subalgebra subregion duality for GRWs. Here we summarize the conclusions for the $\mathcal{Y}_{\hat{R}}$ algebra of the causal wedge and the time band obtained in \cite{Leutheusser:2022bgi}. Consider a time band on the boundary $T_{p,f}$ bounded by $\Sigma_p$ to the past and $\Sigma_f$ to the future. A time band is a causally incomplete boundary spacetime subregion, but it could still have an associated subalgebra. The algebra in $T_{p,f}$ is $\mathcal{Y}_{T_{p,f}}$, i.e. $(\mathcal{M}_{T_{p,f}})''$. By decomposing the boundary time band $T_{p,f}$ into the union of an infinite number of (overlapping) causal diamonds ${D_i}$, with each causal diamond's algebra $\mathcal{M}_{D_i}$ equal to the bulk algebra in the corresponding causal wedge of the boundary  causal diamond $\mathcal{M}_{D_i}=\Tilde{M}_{\mathcal{C}_{D_i}}$,  it could be shown that the subalgebra in the time band $(\mathcal{M}_{T_{p,f}})''$ is equal to the bulk subalgebra of the hole spacetime $\tilde{M}_{\mathcal{C}_{T_{p,f}}}$ \cite{Leutheusser:2022bgi}
\begin{equation}
(\mathcal{M}_{T_{p,f}})''=\vee_{i}\mathcal{M}_{D_i}=\vee_{i}\Tilde{M}_{\mathcal{C}_{D_i}}=\Tilde{M}_{(\cup_{i}\mathcal{C}_{D_i})''}=\tilde{M}_{\mathcal{C}_{T_{p,f}}}
\end{equation}
from the additivity of the two algebras.

Now this process to get the final equivalence of the boundary subalgebra and the bulk subalgebra in the dual subregion could be generalized to the case of  GRWs that only cover part of the boundary spatial subregion easily. The boundary spacetime subregion is now a pancake-shaped region with a finite-sized spatial subregion instead of a time band. Similarly, a pancake-shaped spacetime subregion could be written as a union of an infinite number of causal diamonds and all the formulas above still apply. Thus for GRWs that only cover part of the boundary spatial subregion, we also have the subalgebra-subregion correspondence, which serves as a first piece of evidence for the GRW spacetime subregion duality.

\subsubsection{Equivalence of the entanglement entropy of the reconstructed state and the GRW area}
\noindent
As we mentioned in section 4.2, the main obstacle to identifying the dual state to the one with removed long-range entanglement is if this procedure could always be done consistently. In this subsection, we first explain this problem and specify the corresponding long-range entanglement structure in more detail, and take the tripartite entanglement systems as examples to show that this procedure could be done for certain states and could not be done for other states. Then we will provide evidence that for holographic states the corresponding long-range entanglement could always be removed consistently following the boundary causal structures.

There have been discussions on whether the long-range entanglement structure for a time band could be removed in the literature, which is based on the concept of differential entropy. Differential entropy is the quantity on the boundary that is dual to the non-extremal surface (hole) area in the bulk \cite{Balasubramanian:2013lsa,Balasubramanian:2018uus,Engelhardt:2018kzk}, named ``hole-ography". The information-theoretic interpretation of differential entropy is discussed explicitly in \cite{Czech_2015}. In the following, for our purpose to connect the differential entropy to the boundary dual state, we will focus on the effect of the ``time cutoff" and provide an interpretation of the differential entropy that is mathematically equivalent but in a different physical perspective to the one presented in \cite{Czech_2015}.

 In \cite{Balasubramanian:2013lsa,Balasubramanian:2018uus}, it is suggested that the differential entropy reveals the entanglement between scales \cite{Balasubramanian:2011wt}, specifically the entanglement between UV/IR degrees of freedom on the boundary divided by the time cutoff. However, this interpretation is somewhat ambiguous because it is difficult to explicitly determine which IR degrees of freedom have been cut due to the energy-time uncertainty of the observer on the boundary, which is constrained to a finite time interval. Later, differential entropy is shown to correspond to the {cost to transmit the state of a subregion under the conditions of constrained state merging}. The open question of whether differential entropy could be a von Neumann entropy is proposed in \cite{Balasubramanian:2018uus}. The challenge of answering it and proving the GRW subspacetime duality lies in mathematically defining the ``time cutoff" on the boundary.

Here, as we already elaborated in section 4.2, we propose another interpretation: the observer could observe all degrees of freedom on the boundary, but the observer cannot detect the long-range entanglement structure due to the causality constraint imposed by the time strip. This interpretation is motivated by the realistic observation process in quantum information theory, where an observer must have causal connections with two separated particles both from the past and in the future in order to test the existence of entanglement between them. Theoretically, the observer needs to send particles and classical signals to prepare quantum operators and receive the results of those operators performed on the two particles, respectively. We do not delve into the specific details of this process in quantum information theory. Instead, we only need to understand that an observer living in the boundary CFT (referring to any boundary observer on a single worldline) could only detect the density matrix inside the spacetime subregion that is causally connected to him/herself, namely the observer's causal diamond.

Could a single observer in a definite spacetime subregion detect the whole density matrix $\rho_A$ of the spacial region $A$ as the EW does? If and only if the subregion $A$ in the CFT is spherical in shape, there exists a single observer whose worldline passes through the two vertices of the causal diamond $D(A)$, and this observer is causally connected with the entire region $A$. As a result, the observer has access to the full information of $\rho_A$. This explains from the observer's perspective why in this case CW is equal to EW. On the contrary, in other kinds of shapes of $A$, due to the presence of a time cutoff on the boundary CFT, no single observer's worldline inside the pancake region could be causally connected with the entire region. For example, we have mentioned that the time pancake from the intersection of the CW of an oval with the AdS boundary is the causal diamond of the oval region. Observers within this causal diamond pancake region could only observe the entanglement structure that is within a distance less than the minor axis of the oval in both directions. The entanglement structure that spans beyond this distance is inaccessible to them.

The problem at hand is how to quantify the ``ignorance" of the boundary observers residing within a particular spacetime subregion, with the hope that it corresponds to the entanglement entropy of the GRW in the bulk. The proposed solution is to introduce a fictitious density matrix $\tilde{\rho}_A$ seen by the boundary observers, instead of considering the real density matrix $\rho_A$. The key feature of $\tilde{\rho}_A$ is that its reduced density matrix in {any local spatial subregion whose causal diamond is fully inside the pancake spacetime subregion} is the same as that of $\rho_A$, however, $\tilde{\rho}_A$ has been modified to exclude certain long-range entanglement pairs within $A$ and instead entangle them with regions outside of $A$. Finally, the ``ignorance" is quantified as the von Neumann entropy of $\tilde{\rho}_A$, naturally.

Let us take the ``zigzag" shape time cutoff as a simple example. As depicted in Figure \ref{bcutoff}, the ``zigzag" time cutoff excludes all entanglement between regions $A$ and $B$ as no single observer could observe any two points in A and B simultaneously while preserving the entanglement structure within regions of $AE$ or $BE$. The amount of entanglement ignored compared to the original state is equal to the conditional mutual information $I(A:B|E)$, which, given the information of E, describes the increasing amount of the information of B when A is known. We could interpret this process as counting the ``out legs" due to the time cutoff, as shown on the right side of Figure \ref{bcutoff}. Therefore, we could envision the total ignorance of the observers within this zigzag region as
\begin{equation}\label{Sign}
    S^{ign}_{ABE}=S_{ABE}+I(A:B|E)=S_{AE}+S_{BE}-S_{E}.
\end{equation}
We could perform this analysis successively with additional regions $C, D, \ldots$ by considering the entire $S^{ign}_{ABE}$ as one subsystem. If we take regions $B$, $C, \ldots$ to be sufficiently small such that we can take the continuous limit, this formula will coincide with the formula for differential entropy \cite{Balasubramanian:2013lsa} thus equal to the area of GRWs.

\begin{figure}[H]
    \centering    \includegraphics[width=0.9\textwidth]{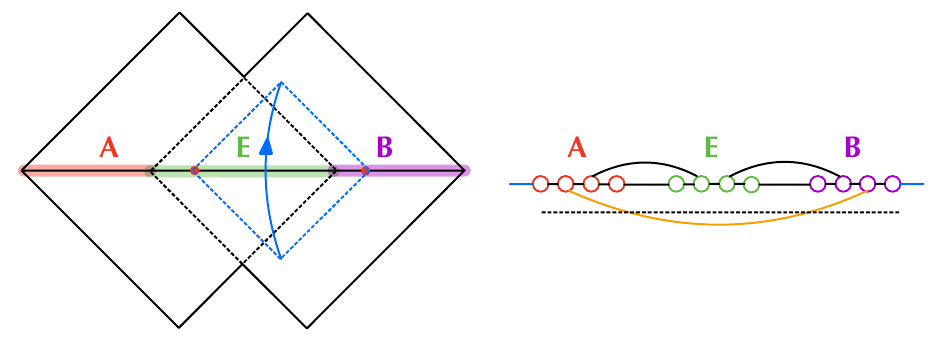} 
    \caption{Left: A `zigzag' shape spacetime subregion with a time cutoff in the boundary spacetime. A boundary observer, whose worldline is the blue curve, could observe the entanglement structure inside a causal diamond marked by dashed blue lines. The observer is free to exist within the zigzag region between the time cutoff. As a result, any entanglement structure inside regions $AE$ or $BE$ could be observed. However, the entanglement structure between regions $A$ and $B$ is beyond observation due to the time cutoff. Right: An illustration of the entanglement structure is shown. Solid lines and curves represent the ``entanglement pairs". The number of blue lines represents the entanglement entropy $S_{ABE}$. The black lines and curves represent the observable entanglement structure, while the orange curve represents the entanglement that cannot be observed due to the time cutoff (represented by the dashed line).} \label{bcutoff}
\end{figure}

Until now, we have described the essential characteristics of the density matrix $\tilde{\rho}_A$, which is associated with a spacetime subregion on the boundary and whose von Neumann entropy corresponds to the entanglement entropy of the GRW. However, an important question remains: does such a $\tilde{\rho}_A$ exist consistently under the principles of quantum information theory?  This has been studied in \cite{Czech_2015} and it was shown that the corresponding density matrix $\tilde{\rho}_{ABE}$ could be constructed from the local ones as $\tilde{\rho}_{ABE} \equiv \exp \left(\log \rho_{AE}+\log \rho_{BE}-\log \rho_{E}\right)$. $\tilde{\rho}_{ABE}$ could be a consistent density matrix only when it is a normalized one, which requires that $\rho_{AE}$, $\rho_{BE}$ and $\rho_{E}$ commute. However, they gave a counterexample from a free conformal field theory showing that this is not always satisfied as the maximal entanglement entropy of $ABE$ is $2/3$ of the differential entropy. As this property apparently depends on the entanglement structure of the system under study and the counterexample does not come from holographic field theories, from the consistency of observers on both sides of holography for GRWs, one would believe that for holographic field theories, we could obtain the $\tilde{\rho}_{ABE}$ consistently. It is important to note that the entanglement entropy calculated in holography using the area of the RT surface is only a first-order approximation, valid in the large $N$ limit. Therefore, the notion of "commute" should be understood in the context of the large $N$ limit, where the $O(N^2)$ order of the commutator should vanish.

Studying this problem directly for field theories is challenging because our understanding of the explicit entanglement structure in strongly coupled field theories is limited. To gain some general insights, we will study this problem, i.e. whether certain long-range entanglement structures could be removed consistently, in the simplest quantum mechanical system with a nontrivial entanglement structure, the three-qubit systems. This analysis will provide us with some preliminary perspectives. Then we will further prove the existence of $\tilde{\rho}_A$, i.e. the consistent removal of corresponding long-range entanglement structures for holographic field theories.

There are only two different types of nonequivalent entanglement structures for tripartite states and we use them as toy examples here: the GHZ type state and the W type state \cite{Bouwmeester_1999,kafatos2013bell,bengtsson2016brief}. For a $3$ qubit quantum system $ABE$, the GHZ state and the W state are defined as 
\begin{equation}
\begin{aligned}
    &|G H Z\rangle_{AEB}=\frac{1}{\sqrt{2}}(|0_A0_E0_B\rangle+|1_A1_E1_B\rangle),\\
    &|W\rangle_{AEB}=\frac{1}{\sqrt{3}}(|0_A0_E1_B\rangle+|0_A1_E0_B\rangle+|1_A0_E0_B\rangle).
\end{aligned}
\end{equation}
As those two states are symmetrical to systems $A,B,E$,  all two partite density matrices of each of the two states are equal
\begin{equation}
\begin{aligned}
&\rho_{GHZ_{AE}}=\frac12(|0_A0_E\rangle\langle0_A0_E|+|1_A1_E\rangle\langle1_A1_E|),\\&\rho_{W_{AE}}=\frac13|0_A0_E\rangle\langle0_A0_E|+\frac23|\Psi_{AE}\rangle\langle\Psi_{AE}|),
\end{aligned}
\end{equation}
where $\Psi_{AE}=\frac{1}{\sqrt{2}}(|0_A0_E\rangle+|1_A1_E\rangle)$ is the Bell state of $AE$. We could see from the above formula that the reduced density matrix for $AB$ is separable for the GHZ state while entangled for the W state. Intuitively, we could view the GHZ state as the state that maximizes the tripartite global entanglement of $ABE$ while the W state as maximizing the bipartite entanglement between $AE, BE$ or $AB$.

Now let us come to the question if the reconstructed density matrix $\tilde{\rho}_{ABE}$ whose von Neumann entropy is equal to ($\ref{Sign}$) exists, which has at the same time the same $AE$, $BE$ reduced density matrices as $\rho_{ABE}$, i.e to consistently remove the entanglement between $A$ and $B$. For the GHZ state, this reconstruction is easy to get
\begin{equation}
\tilde{\rho}_{ABE}=\frac12(|0_A0_E0_B\rangle\langle0_A0_E0_B|+|1_A1_E1_B\rangle\langle1_A1_E1_B|).
\end{equation}
One could calculate the von Neumann entropy of $\tilde{\rho}_{ABE}$ and check that (\ref{Sign}) is indeed the required reconstructed density matrix with the entanglement structure between $A$ and $B$ removed.
However, it could be proved that no such reconstruction could be found for the W state. This is due to the fact pointed out in \cite{Czech_2015} that $\rho_{W_{AE}}$ and $\rho_{W_{BE}}$ do not commute with each other. 
We could also understand this result intuitively as the entanglement of $AB$ is associated with the entanglement of $AE$ and $BE$ in a certain way so that given the latter, the entanglement of $AB$ in the W state cannot be broken.

However, as we already argued above, in holographic field theories \footnote{We have not introduced matter fields to calculate higher-order quantum corrections in the bulk. As a result, our focus is primarily on cutting off the conditional mutual information between $A$ and $B$ in a semiclassical sense. This means we aim to make it vanish at order $O(\frac{1}{G_N})$ in the bulk and $O(N^2)$ on the boundary.}, entanglement structures associated with Rindler convex bulk subregions should not have a W state type.  
To guarantee the existence of $\tilde{\rho}$, we need to construct it in a holographic sense. As long as we could construct a state whose entanglement entropy is exactly the differential entropy, the reconstructed state is
the $\tilde{\rho}$ that we require because $\tilde{\rho}$ should be the maximum entropy state as argued in \cite{Czech_2015}. Now we should proceed to show explicitly that a boundary subregion with time cutoff as exactly indicated by the GRW boundary time range has an entanglement entropy equal to the differential entropy, i.e. the area of the GRW. The simplest case would be to consider a uniform time cutoff that does not depend on the coordinate and all the calculations should be in principle generalized to more complicatetd cases. In this case, an elegant way to get the entanglement entropy of the boundary subreion is to use the RT formula in a bulk geometry whose corresponding boundary theory has a uniform time cutoff, i.e. the hole spacetime. In this case, it couuld be shown that for a given small boundary subregion, the entanglement entropy obtained from the RT formula is exactly the surface area of the GRW for this small subreion in the orginal spacetimm without a hole. The details of this calculation will be shown in the last subsecetion of the next section, as this requires some more techinal background.

\section{Time/space cutoff in the bulk and bulk-boundary observer concordance}
\noindent
As we already mentioned in previous sections, the entanglement wedge is in general not a GRW (except when EW=CW), and this means that in the bulk the entanglement wedge corresponds to a set of non-physical observers. An immediate question is if we could connect the entanglement wedge formalism with the GRWs. Could we turn an entanglement wedge into a GRW? Could an EW also be a gravitational subregion that cannot be observed by a certain set of physical observers? As minimal surfaces in general are not Rindler convex meaning that the observers inside EW always collide at some spacetime points. Thus there are two different ways to make an EW become a GRW: introducing a time cutoff in the bulk spacetime and introducing a space cutoff in the bulk spacetime. 
The former (time cutoff) relaxes the Rindler-convexity condition in the bulk as previous colliding observers do not collide now because the time range for their existence is shorter, thus enlarging the set of GRWs so that a minimal surface might become Rindler-convex now. On the other hand, the latter (space cutoff) could lead to different RT surfaces to make them coincide with Rindler-convex surfaces. These two perspectives also help us gain a deeper understanding of the spacetime subregion duality. The space cutoff method (section 5.2) gives us supporting evidence in proving the GRW spacetime subregion duality as mentioned at the end of the last section. The time cutoff method indicates a further generalization of the spacetime subregion duality to a proposal of observer concordance, as we will explain in section 5.1.
\subsection{Time cutoff}
\noindent
In the previous section, we provided a detailed explanation of the consequences when a time cutoff is introduced on the boundary. We used observer physics to demonstrate that the time cutoff selectively ``cuts" long-range entanglement while preserving the short-range entanglement structure. This observer physics is expected to be applicable in a general sense and not specific to any particular theory. Building upon this understanding, a natural next step is to extend this concept to the gravitational system and analyze the entanglement structure within that framework\footnote{This approach is discussed more extensively in \cite{Ju:2023dzo}.}.

What could be the effects caused by the introduction of a time cutoff in the gravitational system? For certain Rindler-concave regions whose observer worldlines intersect at a certain point and time, if we introduce a time cutoff so that the intersection point does not belong to this spacetime, then these Rindler-concave regions become Rindler-convex. Briefly speaking, the introduction of a time cutoff in gravitational systems leads to the relaxation of the Rindler convexity condition. This basic observation forms the starting point for further analysis, where we will explore the effects of the time cutoff in both gravitational and holographic theories below.

\subsubsection{Mutual information under time cutoff}
\noindent
In this subsection
, we demonstrate that the time cutoff introduced in the bulk also serves to ``cut" the long-range entanglement, the same as the phenomenon observed on the boundary CFT.

According to our observer interpretation, the entanglement structure between regions $A$ and $B$ is removed due to the inability of observers to exchange entangling pairs or communicate signals between these regions within a finite and short enough time interval. This effectively renders the entanglement structure of $A$ and $B$ unobservable. From a geometric perspective, this reveals that regions $A$ and $B$ do not have a causal connection. In other words, the normal null geodesics emitted from the surfaces of $A$ and $B$ never intersect, either in the future or in the past within the time interval under consideration. If we add the condition that both region $A$ and region $B$ are Rindler convex regions, an interesting observation can be made when viewing $A$ and $B$ as a whole: any normal null geodesics emitted from the boundary of the combined region $AB$ never intersect with each other.  According to the normal condition of Rindler-convexity, the combination of regions $A$ and $B$ itself becomes Rindler convex.

Typically, without time cutoff, Rindler-convex regions are always topologically trivial because if it has two boundaries, the normal null geodesics emitted from those boundaries will intersect with each other. However, the introduction of a time cutoff relaxes the condition of Rindler-convexity, allowing the disconnected region $AB$ to become Rindler-convex with an entanglement entropy of 
\begin{equation}
    S_{AB} = \text{Area}(\partial(A+B)) =\text{Area}(\partial A)+\text{Area}(\partial B)= S_A + S_B
\end{equation}
As a result, the mutual information between regions $A$ and $B$ vanishes, indicating that the entanglement between them is indeed "cut" by introducing this time cutoff, where the entire system can be described by a simple separable state
\begin{equation}
    \rho_{AB}=\rho_A\otimes\rho_B.
\end{equation}

\subsubsection{Holographic observer in time cutoff geometries}

\noindent
With time cutoff in the bulk spacetime it is not known if a well-defined holographic dual exists in general. However, some specific time cutoff introduced in the bulk might have a well-defined holographic correspondence, then in these cases, we could try to analyze the question of whether EW could become a GRW as the Rindler convexity condition is relaxed, this will also lead us to a principle of holographic observer concordance, whose exact meaning will be introduced later.

    The Poincaré coordinate of AdS can serve as an interesting and important example. The Poincaré patch can be seen as the global AdS spacetime with a null-shaped time cutoff, as depicted in Figure \ref{Poincarépatch}. In the original spacetime, Rindler-convexity is equivalent to geodesic convexity. However, after introducing this cutoff, certain regions that were originally geodesic concave can become Rindler convex now. We could analyze the Rindler convexity condition with the time cutoff using the tangential condition introduced in Section 2. 
    
    By definition, the largest lightsphere is the intersection of the light cone and the Cauchy slice, with the vertex of the light cone placed on the boundary of the spacetime. In the case of the Poincaré patch, there are two classes of surfaces that can be considered as the largest lightspheres due to the presence of two boundaries. The first class consists of planes, which are the intersection of the Cauchy slice and the light cone with the vertex on the AdS boundary. The second class consists of ``hyperspheres" and ``horospheres", which are the intersection of the Cauchy slice and the light cone with the vertex on the time cutoff surface. It is worth noting that any lightsphere in the second class must intersect at the same point in global AdS, which in the Poincaré patch is the point at infinity. Consequently, these lightspheres correspond to ``hyperspheres" and ``horospheres", which are well-defined in hyperbolic space within classical differential geometry. 
    
    In the Poincaré coordinate, the analysis becomes simpler as the metric is static and the null geodesics are straight lines. The two classes of the largest lightspheres mentioned earlier correspond to semicircles and straight lines, respectively. The question arises of when to use the first class or the second class when analyzing the Rindler convexity of a given subregion. The key factor is the concept of being ``\textbf{externally} tangential" in the tangential condition. The inner side of a lightsphere is the inner side of the light cone. In global pure AdS, both sides of a plane could be the inner side, while in the case of the Poincaré patch, the inner side of a semicircle must be the region between the semicircle and the AdS boundary, while the inner side of a straight line is the ``larger" region where the angle between the straight line and the AdS boundary is greater than $\frac{\pi}{2}$. Only one of those classes can be \textbf{externally} tangential to a given region at most.
    \begin{figure}[H]
        \centering 
        \includegraphics[width=0.6\textwidth]{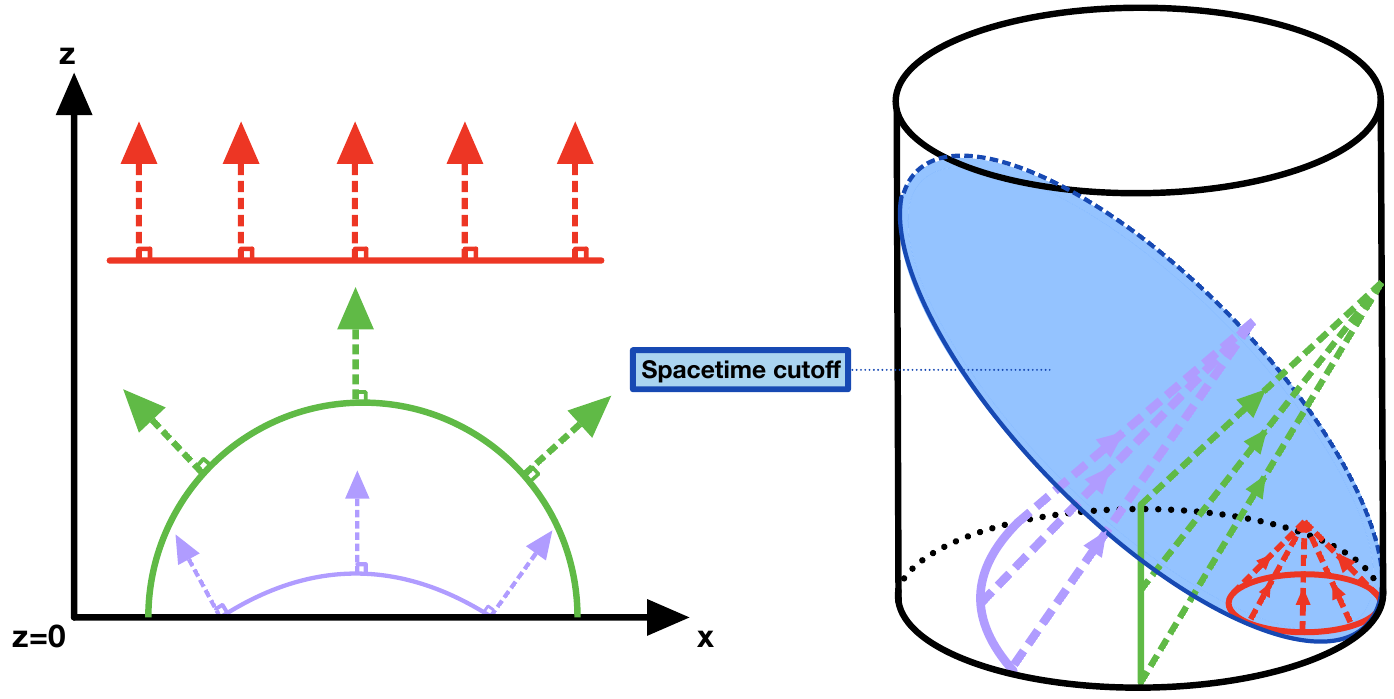} 
        \caption{Left: static Cauchy slice in Poincar\'e coordinate. Right: $t_{global}>0$ half of Poincar\'e patch from global AdS with a null cutoff (blue surface). The red and purple surfaces towards the $z>0$ side are Rindler-convex in Poincar\'e patch while Rindler-concave in global AdS.}
        \label{Poincarépatch} 
    \end{figure}
    
With a complete understanding of the Rindler-convexity condition after the time cutoff in the Poincar\'e patch, the next step is to analyze the RT surfaces and determine which types of boundary subregions become Rindler convex in this scenario. As a reminder, only the RT surface of spherical regions on the boundary are Rindler convex in global AdS. Now, we propose a proposition here: the RT surface of a boundary subregion $A$ is Rindler convex in Poincaré patch if $A$ is a convex set on the boundary.

Geometrically proving this theorem is challenging due to the difficulty in analyzing the global characteristics of minimal surfaces. However, there are strong pieces of evidence supporting this result, which we will briefly mention here:
1. Numerical calculations have been performed on various RT surfaces of convex sets, and the results consistently indicate the Rindler convexity of the EW.
2. The RT surface lies inside the causal shadow of the boundary region $A$, and the boundaries of the causal shadow, $\partial CW(A)$ and $\partial CW(A^c)$, are both Rindler convex if and only if $A$ is Rindler convex.
3. An explicit analysis of the partial differential equations that govern minimal surfaces rules out the existence of saddle-shaped RT surfaces in Poincaré patch which is Rindler-concave for convex regions $A$, providing further evidence for the Rindler convexity of the EW.
These pieces of evidence, while not constituting a rigorous geometric proof, strongly suggest the validity of the inverse version of the theorem. Building on the assumption that this theorem holds, we will further explore its implications and consequences.

When we say that $EW(A)$ is Rindler convex, it implies that $EW(A^c)$ corresponds to the GRW. However, $EW(A^c)$ is not necessarily Rindler convex, except in the case when $A$ is a spherical region on the boundary. This is because $A^c$ is concave on the boundary. The distinction between convexity and concavity is a matter of direction. For a given surface, it can be convex on one side and concave on the other side. In practice, when testing the Rindler convexity of $EW(A)$, we use the second class of the largest lightspheres which include the point at infinity. Conversely, when testing the Rindler convexity of $EW(A^c)$, we use the first class of the lightspheres. This difference arises because only the ``outgoing" direction of Rindler convexity is relaxed in the Poincaré patch compared to the global AdS.

On the flat boundary of the CFT, we define a set $A$ as an ``observable region" if the reduced density matrix $\rho_A$ is detectable by consistent physical observers. In this sense, any set $A$ whose complement $A^c$ is convex can be considered observable because the Rindler observers of $A^c$ can detect $\rho_A$. Finite-size Rindler convex sets also fall under this category. For example, observers inside a spherical region $A$ can observe the reduced density matrix $\rho_A$, making $A$ an observable region.

What about the combination of two spherical regions $A$ and $B$? Typically, one would not consider it as an observable region because even though the observers inside $A$ and $B$ could detect $\rho_A$ and $\rho_B$ respectively, the combined set of observers usually cannot detect $\rho_{AB}$ due to the non-zero mutual information between $A$ and $B$. However, when $A$ and $B$ are far apart, the RT surface of $AB$ becomes disconnected, the mutual information vanishes so that $\rho_{AB}$ can be expressed as the tensor product of $\rho_A$ and $\rho_B$. In this case, one can consider $\rho_{AB}$ as observable. 

By considering the spherical regions and the combination of distant spherical regions, we can further generalize the proposition as follows: $EW(A)$ is a GRW in the bulk if and only if $A$ is an observable region on the boundary\footnote{In a finite-sized CFT that corresponds to global AdS, the observable regions are limited to spherical regions and combinations of distant spherical regions. Their EWs are also GRWs in the bulk.}. This generalization provides a holographic correspondence for the observations of observers.

As we delve into the understanding that observers play a crucial role in partitioning the degrees of freedom within a gravitational system, we have come to realize that this partitioning has a holographic nature, which we refer to as the ``observer concordance". For an observable region $A$ on the boundary, its dual subregion in the bulk ($EW(A)$) must be a GRW that can be partitioned and observed by bulk Rindler observers. For the unobservable regions on the boundary where physical observers cannot observe the entire entanglement structure inside $A$ without being causally connected with $A^c$, we introduce a time cutoff and construct $\tilde{\rho}_A$ as the density matrix detected by observers. This construction corresponds to the GRW in the bulk, as indicated by the GRW subspacetime duality. From this perspective, we can incorporate the GRW subspacetime duality into the holographic observer concordance framework：GRW duality gives a holographic correspondence from GRW in the bulk to the $\tilde\rho$ on the boundary, \ie{} $\text{bulk observers}\to \text{boundary observers}$; now we reverse the dual and say that the EW of any observable region on the boundary must be a GRW in the bulk, \ie{} $\text{boundary observers}\to \text{bulk observers}$ and the holographic observer concordance refers to the combination of both the two directions.

\subsubsection{A volume law}
\noindent
For a boundary subregion $A$ whose characteristic length is $R$, we introduce a time cutoff with a thin time layer $-t_c/2 < t < t_c/2$ left on the boundary, where $\epsilon\ll t_c\ll R$, $\epsilon$ the UV cutoff on the boundary. The dual bulk GRW is located very close to the boundary, as shown in Figure \ref{GRWUVIR}. The entanglement entropy of this GRW will be proportional to the volume of the boundary subregion
\begin{equation}\label{CFTvolume}
    S_{t_c}(A)=Area(GRW)\sim a_{UV} \frac {Area(\partial A)}{\epsilon^{d-2}}+\frac{Vol(A)}{4G_N(t_c/2)^{d-1}}. 
\end{equation}
The first term represents the usual UV divergent term caused by the entanglement pairs near $\partial A$ and $a_{UV}$ is a constant \cite{Rangamani_2017}. After introducing the time cutoff, only the short-range entanglement remains. {Long range entanglement are all ``cut" and contribute to the total entanglement entropy} with its amount proportional to the volume of $A$ as indicated by the second term in \ref{CFTvolume}. This configuration aligns with our intuition that long-range entanglement purifies the short-range entanglement, which is proportional to the volume, and ultimately only the short-range entanglement across the entangling surface 
remains, resulting in the entanglement entropy being proportional to the area of the entangling surface.

The discussion above is for the case when the boundary spacetime subregion has a large time cutoff. As we are discussing the physics related to the time cutoff in the bulk in this section, we will now analyze whether similar consequences arise in the gravitational system with a bulk time cutoff. In other words, does the volume law also appear in the gravitational system when the spacetime has a large time cutoff? This presents an opportunity for us to gain more understanding into the gravitational entanglement structure.

Motivated by this, we introduce a time cutoff that leaves an equal-thickness time layer after the cutoff. In this scenario, the largest lightspheres take the form of spheres with a constant radius of $r=t_c/2$ in conformal coordinates. Using these spheres externally tangential to a given region, we can test its Rindler-convexity. Specifically, we are interested in analyzing the case when $t_p\ll t_c\ll t_R$, where $t_p$ is the Planck time and $t_R$ is the characteristic length of a finite-size subregion in the gravitational system that we are studying.

After introducing this time cutoff, Rindler convexity is significantly relaxed and the surface area condition is violated. Consequently, for a finite-size Rindler convex region $A$ in the gravitational system, its entropy may be smaller than the entropy of its subsystem $a$. We can now calculate the difference between $S_A$ and the largest entropy of $a$, denoted as $(S_a)_{\text{max}}$, to reveal the extent to which the degrees of freedom inside $A$ could be further partitioned under this cutoff.

Geometrically, we can imagine filling the finite region $A$ with spheres of radius $r=t_c$ (similar to small bubbles in a water tank). {Imagine that the tank is region $A$ in the gravitational system. Now we want to find its subregion with the largest entanglement entropy which we believe, obeys the volume law. Imagine the water is full of the tank initially, now we add bubbles inside the tank with air inside the bubbles as many as we can, air bubbles occupy as much space as they could. Now, the left water in the tank is a wiggly-shaped item. This wiggly-shaped region is Rindler convex as a result of the fact that Rindler convexity is relaxed. Furthermore, this wiggly-shaped region has the largest surface area, \ie{} the largest entanglement entropy.} In this construction, the Rindler-convex subregion $a$ with the largest entanglement entropy corresponds to the portion of $A$ that is not occupied by these spheres. The difference between $(S_a)_{\text{max}}$ and $S_A$ will be equal to the sum of the surface areas of these ``bubbles".
As the size of the spheres (or ``bubbles") is small enough, the number of bubbles $N$ needed to fill region $A$ is proportional to the volume of $A$ $N\propto Vol(A)/{(\Delta t)}^{d}$ so we have
\begin{equation}
    (S_a)_{\text{max}}-S_A\sim N\frac{\Omega^{d-1}{(\Delta t)}^{d-1}}{4G_N}=  k\frac{Vol(A)}{4G_N\Delta t},
\end{equation}  
where $\Omega_{d-1}{(\Delta t)}^{d-1}$ is the surface area of a bubble in $d$ dimensional flat space. The coefficient of proportionality $k$ in this equation is
\begin{equation}
    k=\frac{\rho\,\Omega_{d-1}}{\Omega_d}=\frac{\rho\Gamma(\frac{d-1}2)}{\sqrt{\pi}\Gamma(\frac{d+1}2)},
\end{equation}
where $\rho$ is the density of the $d-1$ sphere packing in $d$ dimensional flat Euclidean space when $t_c$ tends to zero as the manifold is locally flat.

Similar to the CFT case, the emergence of the volume law indicates that entanglement pairs within the region purify each other, ultimately leaving only the entanglement across the boundary to contribute to the entanglement entropy, with the number of entanglement pairs being proportional to the area in the gravitational system. However, there is a difference due to the presence of the $\Delta t$ term, which reflects the discrepancy in the {behavior of the} density of long-range entanglement between {gravity theory in the bulk and the field theory on the boundary}. In the gravitational system, we cannot guarantee the existence of long-range entanglement, as short-range entanglement can also purify each other locally. For a more detailed discussion of the gravitational entanglement structure, please refer to our future work \cite{Ju:2024xcn}.

    \subsection{Space cutoff}
    \noindent
    Introducing a time cutoff in the bulk, as we discussed earlier, has its limitations. One of the challenges is that, except for some special examples like the Poincaré patch, it is often difficult to establish a convincing holographic correspondence between the bulk time cutoff and its manifestation on the boundary. As a result, in this section, instead of modifying the Rindler-convexity condition to make a RT surface Rindler-convex, we explore the possibility of modifying the geometry inside the bulk to deform the RT surface itself. The goal is to bring the EW closer to, or even coincide with, the GRW.

    This section will also introduce the background knowledge and finish our goal to provide a holographic proof of the GRW subspacetime duality in section 4.2.2. Instead of directly proving the existence of $\tilde\rho_A$ in the large $N$ limit of CFT, our approach is to establish a holographic correspondence for $\tilde\rho_A$. In this new bulk-boundary correspondence, we expect that the entanglement entropy $S(\tilde\rho_A)$, which corresponds to the area of the EW in the new bulk, is equal to the entanglement entropy of the GRW in the old bulk. 

\subsubsection{Modifying the IR geometry}
\noindent
    Let us begin by reviewing the desired properties of $\tilde\rho_A$. The time cutoff on the boundary effectively cuts off long-range entanglement while preserving the short-range entanglement structure. Taking inspiration from the MERA holographic correspondence, where bulk geometry reflects the entanglement structure of boundary states, we can use this perspective to modify the IR geometry of the bulk based on the global AdS. Specifically, since $\tilde{\rho}_A$ and $\rho_A$ have the same reduced density matrix $\rho_a$, where $a$ is a spherical subregion with $D(A)$ located within the time pancake on the boundary, we should ensure that the bulk geometry inside $EW(a)$ is preserved. By combining the EWs of all these spherical subregions $a$, we obtain the GRW in the bulk. Therefore, we should refrain from deforming the geometry inside the GRW.

    The second characteristic of $\tilde{\rho}_A$ is that the conditional mutual information $I(A:B|E)$ vanishes, as depicted in Figure \ref{bcutoff}. As shown in Figure \ref{IRreplace}, when we visualize the entanglement wedges $EW(AE)$, $EW(BE)$, $EW(E)$, and $EW(ABE)$ together, we observe that $EW(ABE)$ is always the largest because of the nesting rule \cite{Bousso:2022hlz}, with its edge located deepest in the bulk. Consequently, $S_{ABE}$ is smaller than $S_{AE} + S_{BE} - S_E$. Given that we may modify the geometry deep inside the bulk, we may be able to find a method to ``repel" $\partial EW(ABE)$ from the IR region of the bulk and bring it closer to $\partial(EW(AE) \cup EW(BE))$. Similarly, $\partial EW(AE)$ and $\partial EW(BE)$ could also be ``repelled" so that $\partial(EW(AE) \cap EW(BE))$ is closer to $\partial EW(E)$, while $\partial EW(E)$ remains unchanged because it is inside the GRW where the geometry cannot be deformed. The closer the edges of entanglement wedges are to each other, the smaller the conditional mutual information $I(A:B|E)$ becomes.
    \begin{figure}[H]
        \centering 
        \includegraphics[width=0.8\textwidth]{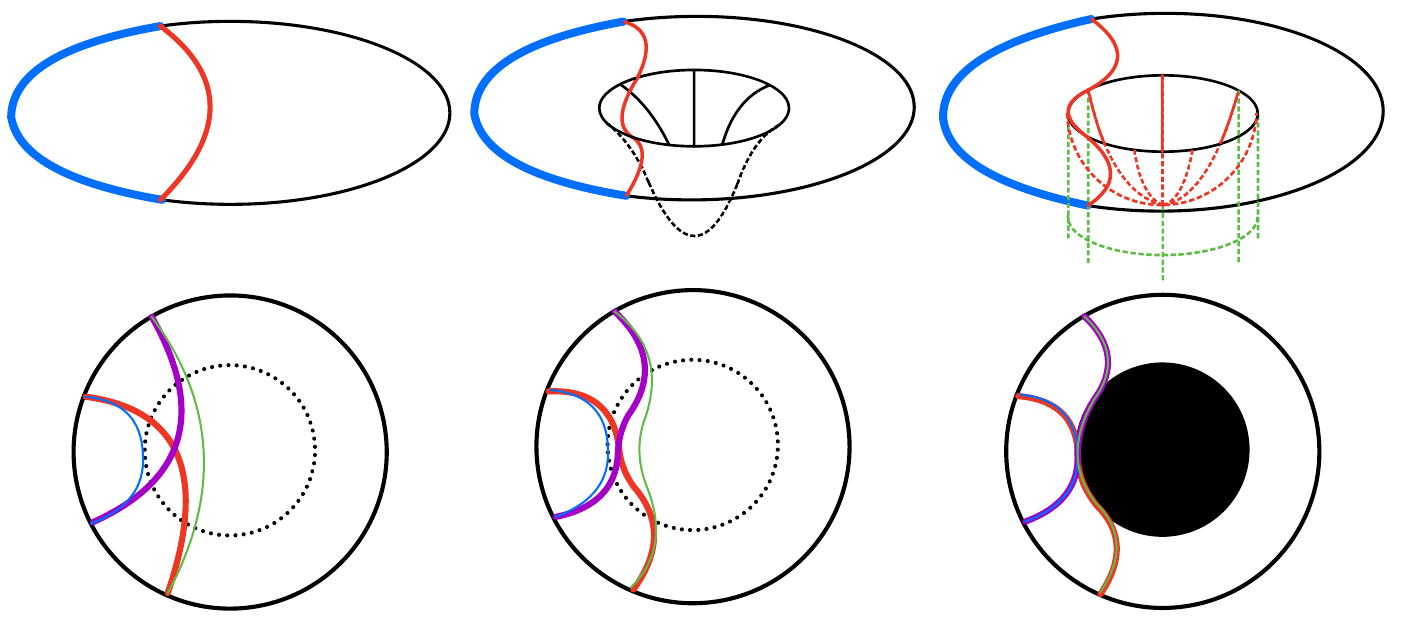} 
        \caption{The $t=0$ Cauchy slice of the modified IR geometry is represented. The figures below show the boundaries of $EW(AE)$, $EW(BE)$, $EW(ABE)$, and $EW(E)$ represented by purple, red, green, and blue curves, respectively. The figures above show Cauchy slices embedded in a high-dimensional fictitious Euclidean geometry. In the left figures, we have the original pure AdS bulk geometry. In the middle figures, the original geometry inside and near the black dashed circle is replaced with a positive curvature space, causing the geodesics to be ``repelled" from the center of AdS and bringing the boundaries of the entanglement wedges closer to each other. In the right figures, we have two extreme cases where RT surfaces are completely repelled out of the IR region. {The difference} between the semisphere (red dashed) and the cylinder (green dashed) IR geometry {appears topologically}. In these cases, the edges of the entanglement wedges coincide with each other, and $I(A:B|E)$ vanishes.}
        \label{IRreplace} 
    \end{figure}

    As depicted in Figure \ref{IRreplace}, we can modify the IR geometry to deform the EW and bring it closer to the GRW. The procedure involves several steps. First, we design a Cauchy slice where the IR geometry is replaced compared to the pure AdS case. Next, we choose a suitable spacetime metric $g_{00}$ in the vicinity of this Cauchy slice, ensuring that the energy-momentum tensor on the Cauchy slice satisfies the null energy condition and that the metric exhibits time reflection symmetry. Finally, we evolve this Cauchy slice both in the future ($t>0$) and past ($t<0$) directions to obtain the geometry of the entire bulk.

    There are a few important points that need to be clarified here. First, when evolving forwards and backward in time, the energy-momentum tensor inside the IR region cannot influence the geometry of the UV region faster than the speed of light. Therefore, the UV geometry remains unchanged, and its domain of dependence, the GRW, remains the same as in the pure AdS geometry. Consequently, $EW(a)$ in the time direction is identical to the pure AdS case, as expected.
    
    Secondly, we should use the HRT formula instead of the RT formula to determine the entanglement wedge, as the spacetime is dynamical. Due to the time reflection symmetry of the Cauchy slice, the minimal surface on it naturally becomes a candidate for the HRT surface. However, is it possible that there exists another minimal surface on a different Cauchy slice with a larger area? The answer is no, and we provide a simple proof. Let us assume there are two candidate HRT surfaces, each being the minimal surface on its own Cauchy slice with zero null extrinsic curvatures. We could ``null project" one of them, denoted as $\text{HRT}_1$, onto the Cauchy slice where the other candidate, $\text{HRT}_2$, lies. We name the surface after the projection as $\text{HRT}_1'$. Under the null energy condition, according to Raychaudhuri's equation, the surface area must decrease in this ``null projection" process, i.e., $Area(\text{HRT}_1) > Area(\text{HRT}_1')$. At the same time, $\text{HRT}_2$ is the minimal surface on Cauchy slice 2, so we have $Area(\text{HRT}_1) > Area(\text{HRT}_1') > Area(\text{HRT}_2)$. Note that we have not used any actual differences between $\text{HRT}_1$ and $\text{HRT}_2$, yet we obtain an asymmetric inequality of their areas. By exchanging the indices, we can derive $Area(\text{HRT}_2) > Area(\text{HRT}_2') > Area(\text{HRT}_1)$ using the same logic. These two results contradict each other. Therefore, we have proven that there cannot exist two candidate HRT surfaces at the same time. Hence, the minimal surface on the time reflection symmetric Cauchy slice must be the unique HRT surface.

    Thirdly, when designing the geometry of the Cauchy slice, various values of the curvature scalars from matter fields could be employed to support the IR geometry. However, they must satisfy certain global conditions governed by the Chern-Gauss-Bonnet theorem. Taking a 2d Cauchy slice in a 3d AdS spacetime as an example, the Gauss-Bonnet theorem relates the integration of the curvature scalar inside the hole to the integration of the extrinsic curvature along the edge of the hole
    \begin{equation}\label{GB}
        \int \sqrt h K + \int k_g ds =2 \pi \chi(M),
    \end{equation}
    where $K$ represents the Gauss curvature on the Cauchy slice, which is half of the curvature scalar, $k_g$ is the geodesic curvature of the edge of the hole, and $\chi(M)$ is the Euler characteristic number associated with the topology of the IR region. Modifying the geometry inside the IR region can be understood as redistributing the curvature scalar within the region. Repelling the RT surface corresponds to increasing the curvature near the center of the hole, resulting in a decrease in curvature near the edge of the hole. As the Cauchy slice has zero extrinsic curvature, the Riemann tensor on this Cauchy slice ($^3R_{abc} {}^d$) is equal to the projection of the Riemann tensor of the spacetime onto this Cauchy slice \cite{Carroll:2004st}
    \begin{equation}
    \begin{aligned}
        &^3R^\rho{}_{\sigma \mu\nu} =P^\rho{}_\alpha  P_\sigma{}^\beta P_\mu{}^\gamma P_\nu{}^\delta R^\alpha{}_{\beta\gamma\delta}-(K^\rho{}_\mu K_{\sigma\nu}-K^\rho{}_\nu K_{\sigma\mu}),\\
        &^3R=R+ 2R_{\mu\nu}n^\mu n^\nu+K^2-K^{\mu\nu}K_{\mu\nu},
    \end{aligned}
    \end{equation}
    where $K_{ab}$ and $n^\mu$ represent the extrinsic curvature and the normal vector of the Cauchy slice, respectively.
    {The Einstein constraint equations are:}
    \begin{equation}
    \begin{aligned}
        ^3R-K^{\mu\nu}K_{\mu\nu}+K^2 &=16\pi G \mu, \\
        \mathcal{D}^\mu K_{\mu \nu}-\mathcal{D}_\nu K  &=8\pi Gj_\nu,
    \end{aligned}
    \end{equation}
    {where $\mu$ and $j_\nu$ are the matter energy and momentum densities, respectively.} The null energy condition, given by $R_{\mu\nu}k^{\mu}k^{\nu}\geq 0$, is independent of the curvature scalar ${}^3R$ on the Cauchy slice. Taking the perfect fluid as an example, the null energy condition stipulates that $\mu+p\geq 0$, while $^3 R= 16\pi G\mu$ can take either positive or negative values. {In other words, the weak energy condition is violated while the NEC holds.} Therefore, it is possible to introduce a valid energy-momentum tensor to modify the IR geometry in principle.

Additionally, it is possible to change the topology of the IR region. One interesting topologically nontrivial case is the presence of a two-sided wormhole within the IR region\footnote{Due to the topological censorship \cite{Galloway:1999br}, this wormhole must not be traversable.}, connecting another asymptotic AdS spacetime with another conformal boundary. In such cases, it is important to note that the state of a single CFT can no longer be considered as a pure state, and the area of the wormhole throat corresponds to the entanglement entropy between the states of the two CFTs (as in ER=EPR). It is also possible to construct more complicated wormholes with highly nontrivial topologies. From the perspective of the boundary CFT, modifying the IR geometry is equivalent to altering the long-distance entanglement structure. The process of ``repelling" the entanglement wedges effectively makes them sparser while creating wormholes establishes entanglement connections with another system.

Finally, it is worth noting that since the generalized Rindler wedges can intersect with the boundary, corresponding to a GRW with infinite volume, it is also possible to modify the geometries inside these regions. Further discussions on this topic can be found in our forthcoming work \cite{Ju:2024xcn}.

\subsubsection{The extreme case: space cutoff}
\noindent
    Let us return to the goal of making the EW a GRW. In the process of repelling the EW, the extreme case is when all geodesics are repelled to the edge of the IR region, as shown in Figure \ref{IRreplace}. One possible way to achieve this is by placing a brane on the edge of the hole and setting the extrinsic curvature from the inside to zero. For example, in the case of a spherical hole, we can imagine replacing the IR geometry with a semisphere (indicated by the red dashed lines in Figure \ref{IRreplace}), where the edge of the hole corresponds to the equator of the semisphere. The curvature singularity on the edge is caused by the tension of the brane. In this scenario, geodesics must circle around the hole and cannot enter it.

    Another option is to create a ``cylinder-shaped" hole (indicated by the green dashed lines) that extends infinitely in depth, effectively changing the topology in the IR region. This cylinder can potentially connect to another bulk region, as mentioned earlier. In the following discussion, we will focus on this case. However, it is important to note that there are other possibilities as well, such as digging a small hole near the center of the semisphere, etc. These geometries correspond to CFT states with richer long-range entanglement structures. Further exploration of these geometries will be discussed in our upcoming paper \cite{Ju:2024xcn}.

    As the entanglement wedge cannot pass through or contain the ``cylinder-shaped" IR region, we can view this situation as simply ``deleting" the region inside a Rindler convex hole. \footnote{This hole, referred to as the region that cannot be penetrated by any boundary-anchored extremal surfaces, is an ``extremal surface barrier" \cite{Engelhardt_2014} and must exist behind a horizon. In our case, it coincides with the horizon. It slightly differs from the concept of the ``entanglement shadow" or ``entanglement hole" \cite{Rangamani_2017, Nogueira_2013, Balasubramanian_2015} because it is excluded by both $EW(A)$ and $EW(A^c)$, where $A$ is a boundary subregion.} According to the surface area condition, there are no other surfaces wrapping this Rindler-convex hole that have a surface area less than the surface area of the hole itself. Therefore, the entanglement entropy of the entire boundary is determined by the area of this Rindler convex hole. This shows that the entanglement entropy of the dual time band state is indeed the area of the hole, giving a fine-grained entropy explanation for the differential entropy.

    To prove the GRW subspacetime duality from the entanglement entropy perspective, it is necessary to first provide rigorous mathematical proof that the idea of ``repelling entanglement wedges could make $I(A:B|E)=0$" is valid, {\ie{}} demonstrating that in the modified geometry, the conditional mutual information $I(A:B|E)$ indeed vanishes. We first look at the shape of entanglement wedges in the modified geometry to study this problem.

    \textbf{The shape of entanglement wedges in modified geometries.} For a small boundary subregion $A$, the modification of the IR geometry does not affect $EW(A)$, which remains the same as in the pure AdS case. However, as we enlarge $A$, a discrepancy with the original $EW$ arises when  $EW(A)$ touches the edge of the Rindler-convex hole, resulting in a second-order phase transition in the entanglement entropy of $A$. After this transition, the RT surface is divided into two parts: the portion outside the hole and the portion coinciding with the edge of the hole. The shape of the entanglement wedge is determined by the partial differential equation that governs these two parts. The former part naturally follows the PDE of a minimal surface in hyperbolic spacetime, while the latter part coincides with the edge of the Rindler-convex hole.

    To determine the complete solution for the entire RT surface, we need to establish the connection conditions between these two parts. It can be demonstrated that these two parts must be smoothly connected. Suppose a sharp ``corner" exists at the junction of the two RT surfaces. In such a case, it is always possible to reduce the area of the RT surface by smoothing out the sharp corners, resulting in a non-minimal surface. To prevent this from occurring, the RT surface outside the hole must be tangential to the hole when it touches the hole, as shown in Figure \ref{EWS}. 
    Consequently, the first part of the RT surface is uniquely determined by the Dirichlet condition on the AdS boundary and the Neumann condition on the surface of the Rindler-convex hole.
    \begin{figure}[H]
        \centering 
        \includegraphics[width=0.7\textwidth]{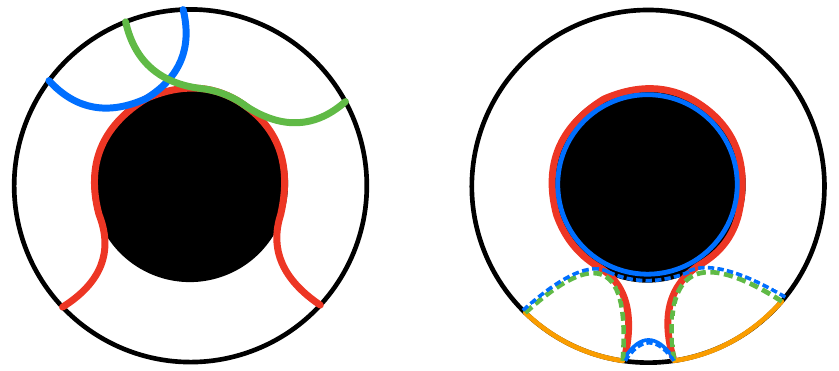} 
        \caption{Left: the figure shows three types of RT surfaces represented by blue, green, and red curves. The black disk represents the Rindler convex hole. For boundary subregions smaller than the one corresponding to the blue (critical) RT surface, the shape of the entanglement wedge remains the same as in the pure AdS case. As the boundary subregion is further enlarged, the entanglement wedges give rise to the green and red RT surfaces, with the two partitions of each RT surface smoothly connected on the edge of the hole. Right: when the boundary subregion becomes sufficiently large, a first-order phase transition occurs in the entanglement entropy. At this point, the RT surface undergoes a transition from the red curve to the blue curve. It is important to note that this phase transition happens at the same time with the phase transition of the shape of the RT surface of the two disconnected boundary subregions marked by orange curves.}
        \label{EWS} 
    \end{figure}
    With this preparation, we can calculate $I(A:B|E)$ holographically. When $EW(E)$ touches the edge of the hole, $EW(ABE)$ is partitioned into three parts like Figure \ref{IRreplace} shows, and we have 
    \begin{equation}
    \begin{aligned}
        &EW(AE)\cup EW(BE)= EW(ABE), ~~~  EW(AE)\cap EW(BE)=EW(E),\\
        &I(A:B|E)=S_{AE}+S_{BE}-S_{E}-S_{ABE}=\text{Area}[\partial (EW(AE)\cup EW(BE))]+\\&\text{Area}[\partial (EW(AE)\cap EW(BE))]-\text{Area}[\partial EW(E)]-\text{Area}[\partial EW(ABE)]=0.
    \end{aligned} 
    \end{equation}
    Finally we have proved that $I(A:B|E)=0$. This has already shown that $S_{ABE}=S_{AE}+S_{BE}-S_E$, meaning that the fine-grained entanglement entropy of a boundary spacetime subregion is equivalent to the differential entropy, i.e. the area of the surface of GRWs.
\subsubsection*{Holographic observer concordance in geometries with IR space cutoff.}

    With the establishment of GRW subspacetime duality, let us now explore the relationship between EWs and GRWs in bulk geometries with the presence of an IR space cutoff. Does the space cutoff result in more EWs transforming into GRWs? This question is crucial, as the holographic observer concordance formalism suggests that EW$=$GRW is a clear indication that the state associated with the boundary spacetime subregion in the sense of long-range entanglement structure removal is physical and could be observed.

    In the case of disjoint observable boundary subregions with connected RT surfaces in pure AdS, \ie each of their own $EW$ is a $GRW$, space cutoff ``repels" the RT surface, causing its area to increase. However, under specific circumstances, it is conceivable that the disconnected minimal surface could have an area smaller than the connected minimal surface after the ``repelling" process, which then leads to an exchange of dominance among these minimal surfaces. As a result, space cutoff causes the EWs of disjointed subregions on the boundary to become disconnected and be a GRW. A connected EW for disjoint subregions is certainly not a GRW. When every disconnected part of the EW is a GRW, the whole is a GRW, which indicates that when each disjoint subregion is observable and there is no entanglement between them, the whole region is observable.

    On the boundary, this process can be understood as follows. Firstly, the space cutoff geometry is dual to the time cutoff on the boundary, meaning that the boundary state has certain long-range entanglement removed. Secondly, as we demonstrated before, the reason why the combination of $n$ disjoint observable regions becomes unobservable in the case without spacetime cutoff is the presence of correlations between these regions. However, with long-range entanglement eliminated due to the space cutoff, the quantum state becomes a direct product state in the semiclassical sense, making it observable again. Thus, the holographic observer concordance framework remains valid, \ie{} the EWs of boundary observable regions  are still bulk GRWs in the modified geometry.

    In fact, the space cutoff does not always make an unobservable region on the boundary observable now; it could also make observable regions unobservable. For a connected boundary subregion $A$, the space cutoff could also make $EW(A)$, which is a GRW in pure AdS, not a GRW now.
    In the following we provide two examples of this possibility.
    
    The first example we demonstrate is the first-order phase transition of the entanglement entropy when the boundary subregion $A$ becomes large enough, as depicted in the right of Figure \ref{EWS}. After this transition, the $EW(A)$ wraps around the Rindler convex hole, and its topology changes. We could see that the boundary of the largest GRW homologous to $A$ is marked by the red curve, and it is smaller than $EW(A)$. The fact that $EW(A) \neq GRW$ implies that the boundary state of the subregion $A$ is no longer a fully observable state/region. In fact, $A$ is indeed no longer an observable state/region because the entanglement structures near the two boundaries of $A$ exist but cannot be observed by observers inside $A$. As depicted in Figure \ref{EWS}, the entanglement entropy of the boundary subregions marked by the yellow curves also undergoes a first-order phase transition at the same time, indicating the presence of entanglement between these yellow regions. However, no boundary observers could observe this entanglement structure within the time pancake of $A$ without being causally connected to its complement, $A^c$. Consequently, $A$ is no longer an observable region on the boundary, which explains why $EW(A)$ is no longer a GRW in the bulk. 

    The second example we demonstrate below, where EW is no longer a GRW after space cutoff, is quite general in high-dimensional $AdS_{d+1}/CFT_d$ where $d\geq3$. Let us take a spherical boundary subregion $A$ as an example. The RT surface homologous to $A$ in pure AdS is a semisphere, which is a unique type of Rindler-convex minimal surface in the bulk. However, as the space cutoff is introduced and modifies $EW(A)$, the (first) part of $EW(A)$ which does not coincide with the edge of the Rindler convex region must be deformed by the Neumann boundary condition on the Rindler convex hole. As a result, the new RT surface must not be Rindler convex. This procedure makes an EW not a GRW after introducing the space cutoff.
    
    As usual, when $EW(A)$ is not a GRW in the bulk, it implies that $A$ is not an observable region on the boundary. This is indeed the case here. As shown in Figure \ref{unob}, on one hand, long-range entanglement ($L>r$) has been eliminated by the space cutoff in the bulk. On the other hand, observers can only observe the entanglement structures within the red circle with radius $r$ inside region $A$. Then, one can see that the entanglement structure marked by the orange lines exists but cannot be observed, which makes the entanglement structure near $\partial A$ unobservable. Back to the bulk, this explains why 
    the part of $EW(A)$ nearby $\partial A$ is Rindler concave. One can imagine that if we ``cut" all the unobservable blue entanglement pairs to make it entangle with the outside, we will find that $A$'s entanglement entropy equals the area of $\partial CW(A)$ shown in green curves. This effect is quite universal in high dimensions, where there always exist unobservable entanglement pairs near $\partial A$ under space cutoff.
    \begin{figure}[H]
        \centering 
        \includegraphics[width=0.95\textwidth]{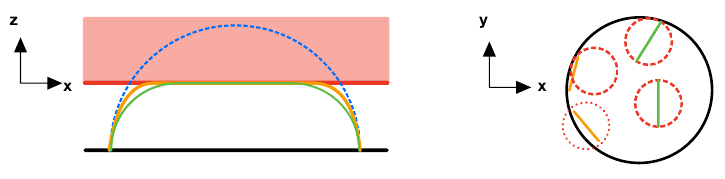} 
        \caption{Left: the EW of the black spherical spatial subregion $A$ in pure AdS is marked by the blue dashed curve, representing a semisphere. The red-shaded region is the Rindler convex region we cut off from the bulk. $EW(A)$ in the bulk after the space cutoff is marked by the orange curve, and its causal wedge is marked by the green curve. $EW(A)$ is not a GRW anymore. The $y$ and $t$ axes are compressed for simplicity.
        Right: on the boundary Cauchy slice, a spherical region $A$ is shown with the $z$ and $t$ axes compressed. Observable entanglement structures are illustrated by red circles totally inside region $A$ while existing bipartite entanglement structures are illustrated by orange and green straight lines. As a result, bipartite entanglement structures marked by green lines are observable, while the orange lines are not, which makes $A$ no longer an observable region after introducing the space cutoff.}
        \label{unob} 
    \end{figure}
    To summarize, the holographic observer concordance framework remains valid under time/space cutoff.
        \textit{The d.o.f in the bulk that can be partitioned by bulk observers correspond to the d.o.f on the boundary that can be partitioned by boundary observers.}

\section{Conclusion and discussion} \noindent
In this work, starting from the physics of accelerating observers in gravitational systems, we show that Rindler-convexity is an important global constraint that plays a crucial role in consistently dividing degrees of freedom in gravitational spacetime into subregions, as well as in the consistency with Page theorem, strong subadditivity, and possibly with the subalgebra in gravitational systems \cite{Leutheusser:2022bgi}. This Rindler-convexity condition is consistent with the convexity condition required in the construction of subregion-subalgebra correspondence \cite{Leutheusser:2022bgi}, which further confirms the consistency of this definition of subregions in gravitational spacetime. 

In holography, we propose new corresponding boundary interpretations of GRWs that only cover part of the boundary spatial subregion, i.e. the GRW/spacetime subregion duality, and compare these to the similar dual interpretations of CWs and EWs.   
We argue that the bulk and boundary geometrical differences reveal the fact that these wedges represent different information in the entanglement structure on the boundary, \ie{} GRW ignores certain long-range entanglement structures due to a limited time range on the boundary. We provide evidence for the GRW spacetime subregion duality from the observer correspondence, subalgebra-subregion duality, and the equivalence of the fine-grained entanglement entropy with the differential entropy, i.e. the GRW surface area. 

We also relax the Rindler convexity by introducing a spacetime cutoff in the bulk so that some EWs become Rindler-convex. A bulk time cutoff could also eliminate long-range entanglement in the gravity subregions. Space cutoff with bulk IR geometry modified could change the long-range entanglement structure of a state and this could precisely give the dual time band state with long-range entanglement structures removed. We emphasize the holographic observer correspondence which states that the bulk subregions that can be partitioned by bulk physical observers correspond to the boundary degrees of freedom that can be partitioned by boundary observers. A volume law in the entanglement entropy for both the boundary theory and bulk gravitational theory exists with a time cutoff in each theory.

{Additionally, an important observation arises within the context of modifying the IR geometry. In this study, we have exclusively explored the scenario where the curvature scalar near the center of the AdS Cauchy slice is increased. However, it is equally possible to modify the geometry in the opposite direction, i.e., by decreasing the curvature. This would lead us to a limit in which the curvature scalar tends towards negative infinity within the IR region. Intriguingly, we have discovered that this bulk geometry is dual to the entanglement structure of the CFT, wherein the long-range entanglement is eliminated and effectively purifies itself without inducing entanglement with other parts of the system. This is very different from the situation in this work. Remarkably, our investigations have revealed that the structure of the RT surfaces precisely aligns with scenarios involving the presence of an end-of-world brane within the bulk. This observation offers a more profound understanding of the AdS/BCFT framework from the perspective of entanglement structures. A comprehensive discussion of these findings will be presented in an upcoming paper \cite{Ju:2024xcn}.}

There are many open questions left. Several of them are the following. Could the generalized entropy formula be obtained from the GRW physics once we introduce matter fields in the gravitational system? Could we confirm the new boundary interpretation of GRW from an explicit bulk reconstruction? Could GRW spacetime subregion duality also exist in flat or de Sitter holography?

\section*{Acknowledgement}
    
We thank Bart{\l}omiej Czech, Yi-hong Gao, Yuan-Chun Jing, Teng-Zhou Lai, Bo-Hao Liu, Jia-Rui Sun, Yu Tian and Xiao-Ning Wu for the useful discussion. This work was supported by Project 12035016 supported by the National Natural Science Foundation of China.

\appendix

\section{Rindler-convexity and geodesic convexity}
\noindent
In this appendix, we prove that in a spacetime with $T_{\mu\nu}k^\mu k^\nu=C_{\rho\mu\nu\sigma}k^\rho k^\sigma=0$ and on any Cauchy slice with zero extrinsic curvature,  Rindler-convexity is equivalent to geodesic convexity, where $C_{\mu\nu\sigma\rho}$ is the Weyl tensor.
Geodesic convexity of region $A$ demands that any plane tangential to region $A$ cannot reach the inside of it \cite{https://doi.org/10.48550/arxiv.1806.06373}, while the tangential condition of Rindler-convexity simply replaces the word `plane' in the geodesic convexity condition with `lightsphere'. As the large lightsphere always contains the small one tangential at the same point due to causality, the tangential condition can be simplified such that the infinitely large lightsphere externally tangential to the boundary of a Rindler-convex region never reaches the inside.
As a result, to prove the equivalence of Rindler-convexity and geodesic convexity under the conditions above, we only need prove that `any infinitely large lightsphere is a plane'.  
We take three steps to prove it as follows.
        
        Step I: we prove that when $T_{\mu\nu}k^\mu k^\nu=0$ and $C_{\rho\mu\nu\sigma}k^\rho k^\sigma=0$, null geodesics emitted from infinity shall never intersect. 
        Raychaudhuri's equations \cite{Hawking:1973uf,Wald:1984rg,Carroll:2004st} of null geodesics are as follows:
        \begin{equation}\label{Raychaudhuri1}
        \begin{aligned}
            \frac{d\theta}{d\lambda}=-\frac1{D-2} \theta^2-\hat\sigma_{\mu\nu}\hat\sigma^{\mu\nu}-8\pi T_{\mu\nu}k^\mu k^\nu+\hat{\omega}_{\mu\nu}\hat{\omega}^{\mu\nu},\\
            k^\rho\nabla_\rho\hat{\sigma}_{\mu\nu}=-\theta \hat{\sigma}_{\mu\nu}+C_{\rho\mu\nu\sigma}k^\rho k^\sigma,
        \end{aligned}
        \end{equation}
        where $\theta$, $\hat\sigma_{\mu\nu}$, and $\omega_{\mu\nu}$ are the expansion, shear, and vorticity of the null geodesics, respectively. The vorticity vanishes for surface-orthogonal null congruences.
        On a congruence of null geodesics locally parallel at some parametrized location $\lambda=\lambda_0$ with $\theta(\lambda_0)=0$ and $\hat\sigma^{\mu\nu}(\lambda_0)=0$, we can solve (\ref{Raychaudhuri1}) when $T_{\mu\nu}k^\mu k^\nu=C_{\rho\mu\nu\sigma}k^\rho k^\sigma=0$, and find $\theta(\lambda)\equiv0, \hat\sigma^{\mu\nu}(\lambda)\equiv0$ for any $\lambda$, which means locally parallel null geodesics stay globally parallel \ie{} they never converge or diverge.
        Thus as we can always construct a globally parallel null geodesic congruence in the neighbourhood of any null geodesic, we can see as if the whole spacetime is `flat' for null geodesics.

        Next, we prove that null geodesics emitted from infinity never intersect. Any two geodesics in a parallel null geodesic congruence have no intersecting point, no matter how close they are. This leads to another way to construct a globally parallel geodesic congruence: we can choose a congruence of null geodesics intersecting at a point in the spacetime and then change the azimuth of each null geodesic in this congruence a little bit in order to push the intersecting point a little bit further. We can use this way to push this intersecting point all the way to null infinity in order to construct a null geodesic congruence without an intersecting point.  
        If another intersecting point emerges inevitably as we move one intersecting point to infinity, we can conclude that no procedure can prevent this null geodesic congruence from intersecting. This conclusion contradicts with the result that we could always construct a parallel geodesic congruence in the neighbourhood of any null geodesic. To this end, we have proved that null geodesics emitted from infinity do not intersect.
        
        Step II: we prove that the null geodesics emitted from infinity must have zero expansion everywhere \ie{} $\theta(\lambda)\equiv0$. 
        We can prove that by contradiction, that is, by assuming $\theta(\lambda_0)=\theta_0\neq 0$, which means $\theta<0$ in a direction. Then we have
        \begin{equation}
            \frac{d\theta}{d\lambda}\leq-\frac1{D-2} \theta^2.
        \end{equation}
        This equation implies that the expansion will become negative infinity $\theta\to -\infty$ (\ie{} intersect) within affine parameter $\Delta\lambda\leq(D-2)/|\theta_0|$ (similiar arguments also appear in \cite{Bousso:1999xy, Hawking:1973uf}). That means null geodesics emitted from infinity intersect, which contradicts with the result in step I.  
        Here we have assumed that the whole spacetime must be geodesically complete so that $\lambda$ could reach $\lambda_0\pm(D-2)/|\theta_0|$, for arbitrary $\theta_0$
        \footnote{As an example, the global AdS satisfies this condition while the AdS Poincar\'e patch does not.}.
        Therefore, the area expansion must vanish on the congruence of the null geodesics emitted from infinity, \ie{} $\theta(\lambda)\equiv0$. 
        
        Step III: we finally prove that the `induced extrinsic curvature' of the infinitely large lightsphere on the Cauchy slice is zero. Once $\theta$ vanishes, $\frac{d\theta}{d\lambda}$ must vanish as well. Combining the Raychaudhuri equation (\ref{Raychaudhuri1}), we get $\theta=\hat\sigma_{\mu\nu}=\hat\omega_{\mu\nu}=0$. Then we have
        \begin{equation}
            \hat B_{\mu\nu}:=\nabla_{\nu} k_{\mu}=\frac12 \theta \hat g_{\mu\nu} +\hat\sigma_{\mu\nu}=0,
        \end{equation}
        where $k_{\mu}$ is the normal null vector, and $\hat{B}_{\mu\nu}$ is known in classical general relativity as the null extrinsic curvature of the null hypersuface  \cite{Bousso:2015mna}.

        We demand that the extrinsic curvature of the Cauchy slice be equal to zero, so we have $\nabla_\nu n_\mu=0$ where $n_\mu$ is the unit normal vector of the Cauchy slice. 
        We can use the linear combination of the vector $k_\mu$ (normal to the null hypersurface) and $n_\mu$ to express the spacial unit normal vector of the Rindler-convex region $l_\mu=\frac {k_\mu}{n_\rho k^\rho}+n_{\mu}$ on the Cauchy slice, then we have
        \begin{equation}            \hat{K}_{\mu\nu}=\nabla_{\nu}l_{\mu}=0, 
        \end{equation}
        where $\hat{K}_{\mu\nu}$ is the induced extrinsic curvature of the lightsphere on the Cauchy slice. At this point, the theorem is proven.

        As an example, in global de Sitter spacetime, it is easy to find out in global de Sitter coordinates that the infinitely large lightsphere is the equator on $t=0$ Cauchy slice. However, the infinitely large lightsphere on $t=constant\neq 0$ Cauchy slice is not the equator because those Cauchy slices do not have zero extrinsic curvature. Moreover, this theorem can also be applied to Minkovski spacetime and global AdS spacetime if we choose a $t=constant$ Cauchy slice whose extrinsic curvature is zero.

        When matter with positive $T_{\mu\nu}k^\mu k^\nu$ (positive matter) exists or $C_{\rho\mu\nu\sigma}k^\rho k^\sigma\neq 0$, null geodesics tend to converge, and Rindler-convexity condition becomes stricter than geodesic convexity on the $K_{\mu\nu}=0$ Cauchy slice.
        Contrarily, when negative matter exists violating NEC or the spacetime is geodesically incomplete, Rindler-convexity may be relaxed and weaker than geodesic convexity on $K_{\mu\nu}=0$ Cauchy slice.
        
\section{Entanglement entropy calculation}
\noindent
To calculate the \ee{} of generalized Rindler wedge by replica trick, we start from ($\ref{metric}$) and perform the Wick rotatation to the Euclidean form:
    \begin{equation}\label{Emetric}
        ds^2=A^2\bigg(\rho^2d\tau_E^2+d\rho^2+[{(\frac{dx_0(\theta)}{d\theta})}^2+{(\frac{dy_0(\theta)}{d\theta})}^2+2\rho \frac{\cos\tau_E}{\cos \theta}\frac{dy_0(\theta)}{d\theta}+\rho^2\cos^2\tau_E]d\theta^2\bigg).
    \end{equation}
    In the near-horizon limit where $\rho$ goes to zero, the metric must be regular. As $\tau$ always enters the metric in $\sinh \tau$ and $\cosh \tau$ terms according to the generalized Rindler coordinate transformation ($\ref{GRT}$), these terms become after Wick rotation $\sin\tau_E$ and $\cos\tau_E$ respectively with an imaginary time period $\beta=2\pi$. 
    In the near-horizon limit which is of our interest, all terms containing $\tau$ vanish because $\rho\to0$. This coincides with the conclusion we highlight in the main text that the horizon is the surface of infinite redshift for the accelerating observer. In general, the near horizon metric behaves
    \begin{equation}\label{ghmetric}
        ds_{nh}^2=A^2|_{\rho\to0}\{\rho^2d\tau_E^2+d\rho^2\}+h_{\theta \phi}dx^{\theta}dx^{\phi},
    \end{equation}
    where $h_{\theta \phi}$ is the induced metric on the horizon.
    The replica trick to calculate the \ee{} is as follows: 
    \begin{equation}\label{replica}
        S=\lim_{n\to 1}\frac{1}{1-n}\log \Tr(\rho_A^n)=\lim_{n\to 1}\frac{1}{1-n}\log (\frac{Z_n}{Z^n}),
    \end{equation}
    where $Z_n$, the  partition function corresponding to the $n$-fold replicated manifold, is to be calculated by the Euclidean path integral.
    The contribution to the partition function except the part from the conical singularity is proportional to the imaginary time period $\beta$, which does not contribute to the \ee{}.
    
    We use Gauss-Bonnet theorem on the $\rho-\tau_E$ surface in order to calculate the contribution from the conical singularity and we have:
    \begin{equation}
        R_n=4\pi (1-n)\delta^2(P) +R_h,
    \end{equation}
    where $\delta^2(P)$ represents the Dirac delta function on the $\rho-\tau_E$ plane, and $R_h$ is the Riemann curvature on the horizon before the replica trick \cite{Fursaev:1995ef,Banados:1993qp}. 

    After that, we can write the partition function $Z_n$:
    \begin{equation}\label{ZN}
        \begin{aligned}
        \log (\frac{Z_n}{Z^n})&= I_{n}-n I=\frac{1}{16\pi G}(\int_{\mathcal{M}_n} \sqrt{g}R_n-n \int_\mathcal{M}\sqrt g R)\\
        & =\frac{1}{16\pi G}\int 4\pi (1-n) d\mathcal{A}\, \int \delta^2 (P)d\tau_E d\rho= (1-n)\frac{\mathcal{A}}{4G}.
        \end{aligned}
    \end{equation}
    Combined with the formula (\ref{replica}), the \ee{} becomes: 
    \begin{equation}\label{SEE}
    S=\frac{\mathcal{A}}{4G},
    \end{equation}
   with $\mathcal{A}$ the area of the Rindler-convex surface. This gives the gravitational entropy of the observer's spacetime, i.e. the complement of the Rindler-convex region, and this is equal to the entanglement entropy between the Rindler-convex region and its complement.
        
\section{Surface area condition under null energy condition}
\noindent
        In gravitational systems, we can imagine if a region of any shape entails an entanglement entropy proportional to its surface area, a region with a smooth boundary surface could have less \ee{} than its subregion with a rougher boundary surface, which violates Page theorem we mentioned above. However, this paradox no longer exists in the context of our work, where only subregions obeying the  Rindler-convexity condition could be associated with a well-defined entropy. Now we prove that under NEC, the surface area of a Rindler-convex region is always larger than its Rindler-convex subregion.
        
        If NEC is satisfied, any Rindler-convex region has a positive expansion along the normal null geodesics pointing outside the region, or the null geodesics will intersect because of the Raychaudhuri's equation as we mentioned in appendix A.
        As we can simply change the signature of the affine parameter $\lambda'=-\lambda$ and $\theta'=-\theta$, any Rindler-convex region has oppositely a negative expansion along the normal null geodesics pointing inside the region. Note that the whole spacetime must be geodesically complete in this argument \cite{galloway1989some}.

        \begin{figure}[H]\label{NECSAC} 
            \centering
            \includegraphics[width=0.4\textwidth]{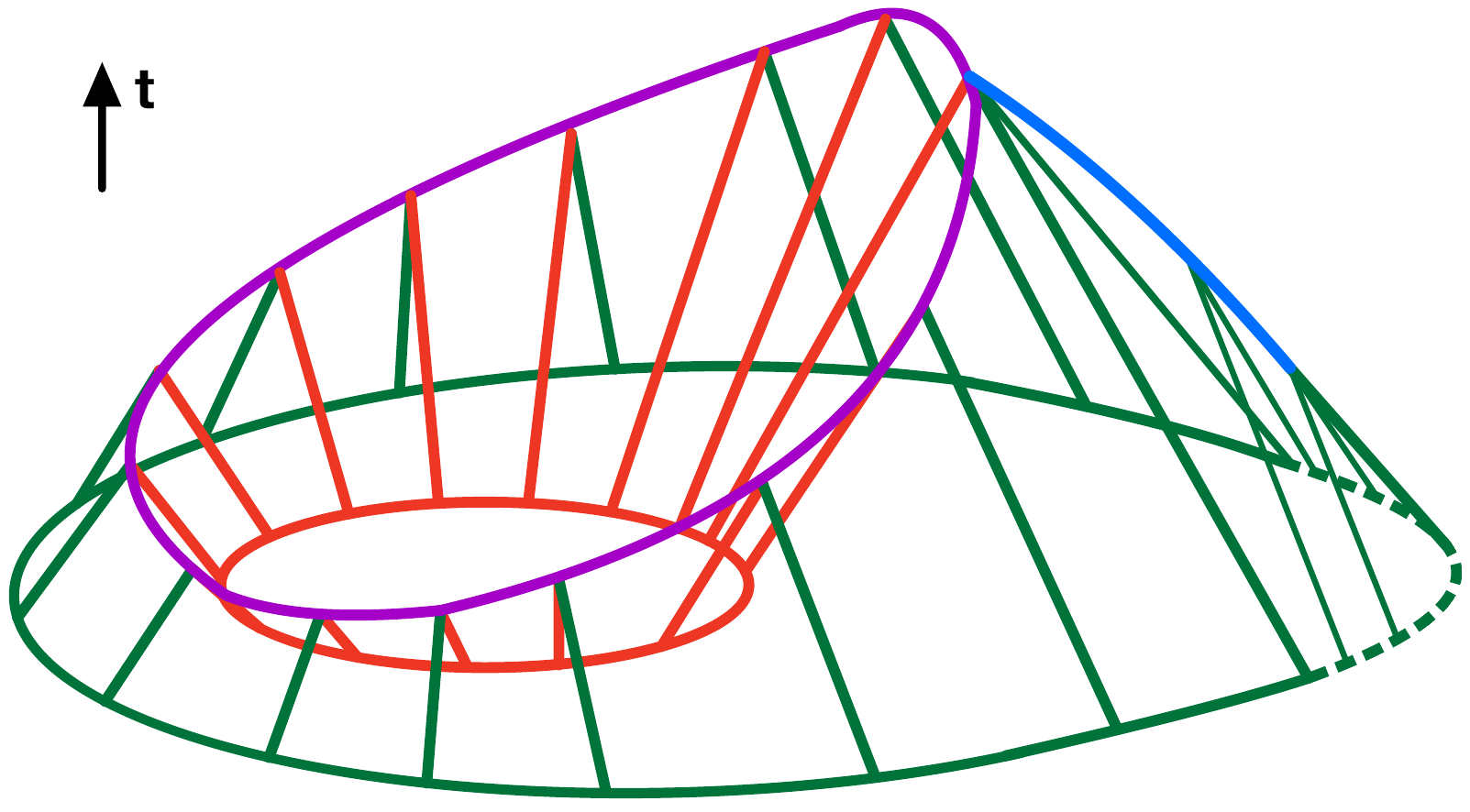} 
            \caption{The green curve is the boundary of the Rindler-convex region while the red closed curve represents its convex subregion. 
            The red straight lines are the normal null geodesics pointing outside the red subregion on a null hypersurface while the green ones are the normal null geodesics going inside the region on a null hypersurface.
            The null geodesics form caustics (marked by the blue line) which constitute a spacelike curve. The intersection of these two hypersurfaces are the spacelike purple line.} 
        \end{figure}
    
        As shown in Figure \ref{NECSAC}, we let the whole region (marked by green oval) `lightlike shrink' and the subregion (marked by red oval) `lightlike expand'. We have:
        \begin{equation}\label{sac}
            Area_{green}>Area_{purple}>Area_{red},
        \end{equation}
        which verifies the surface area condition\@.
        
        Finally, there are some interpretations to highlight. We only demand two conditions to derive surface area condition in this section: 1. NEC 2. 
        Geodesic completeness. 

        Otherwise, Rindler-convexity might be relaxed as a result of the space/time cutoff or the existence of `negative matter'. In this case, Rindler-convexity is weaker than geodesic convexity. As the plane is a trivial kind of minimal surface, we could always build a geodesic-concave but Rindler-convex surface whose boundary is homologous to a plane, which further violates surface area condition.



\bibliography{reference}
\bibliographystyle{JHEP}

\end{document}